\documentclass[12pt,journal,draftcls,letterpaper,onecolumn]{IEEEtran}

\usepackage{dsfont}
\usepackage{amsmath}
\usepackage{amssymb}
\usepackage{amsfonts}
\usepackage{graphicx}
\usepackage{amsmath}
\usepackage[all,poly]{xy}
\usepackage[noadjust]{cite}
\usepackage{multirow}

\newcommand{\figwidth}{0.75\columnwidth}
\newcommand{\figwidthbis}{\textwidth}

\def\bfe{{\mathbf{e}}}

\def\bfs{{\mathbf{s}}}

\def\bfu{{\mathbf{u}}}
\def\bfv{{\mathbf{v}}}

\def\bfx{{\mathbf{x}}}
\def\bfy{{\mathbf{y}}}

\def\bfD{{\mathbf{D}}}
\def\bfE{{\mathbf{E}}}

\def\bfP{{\mathbf{P}}}

\def\bfU{{\mathbf{U}}}
\def\bfV{{\mathbf{V}}}
\def\bfW{{\mathbf{W}}}
\def\bfX{{\mathbf{X}}}

\def\calU{{\mathcal{U}}}

\def\calV{{\mathcal{V}}}

\def\calA{{\mathcal{A}}}

\def\calD{{\mathcal{D}}}

\def\calN{{\mathcal{N}}}

\def\calU{{\mathcal{U}}}
\def\calV{{\mathcal{V}}}

\newcommand{\Vpix}[1]{\mathbf{y}_{#1}}

\newcommand{\MATpix}{\mathbf{Y}}
\newcommand{\pix}[2]{y_{#1,#2}}

\newcommand{\nbpix}{P}
\newcommand{\nopix}{p}

\newcommand{\nbband}{L}
\newcommand{\nbpband}{K}
\newcommand{\noband}{l}
\newcommand{\nopband}{k}

\newcommand{\nbmat}{R}
\newcommand{\nomat}{r}

\newcommand{\MATmat}{{\mathbf M}}
\newcommand{\Vmat}[1]{{\mathbf m}_{#1}}
\newcommand{\mat}[2]{m_{#1,#2}}

\newcommand{\MATpmat}{{\mathbf T}}
\newcommand{\Vpmat}[1]{{\mathbf t}_{#1}}
\newcommand{\pmat}[2]{t_{#1,#2}}
\newcommand{\sampleVpmat}[2]{{\mathbf t}_{#1}^{(#2)}}

\newcommand{\pmatvar}[1]{s^2_{#1}}
\newcommand{\Vpmatvar}{\bfs^2}

\newcommand{\abond}[2]{a_{#1,#2}}
\newcommand{\Vabond}[1]{{\bold a}_{#1}}
\newcommand{\sampleVabond}[2]{{\bold a}_{#1}^{(#2)}}
\newcommand{\MATabond}{{\bold A}}
\newcommand{\Vcoeff}[1]{{\bold c}_{#1}}

\newcommand{\MATcoeff}{{\bold C}}
\newcommand{\Vnoise}[1]{{\mathbf n}_{#1}}
\newcommand{\noise}[2]{n_{#1,#2}}
\newcommand{\MATnoise}{{\bold N}}

\newcommand{\MATnoisevar}{{\boldsymbol \Sigma}_{\textrm{n}}}
\newcommand{\noisevar}{\sigma^2}

\newcommand{\paramvect}{\boldsymbol{\theta}}

\newcommand{\Simplex}{\mathcal{S}}

\newcommand{\Tspace}[1]{\mathcal{T}_{#1}}

\newcommand{\R}{\mathds{R}}

\newcommand{\inv}{^{-1}}

\newcommand{\transp}{^T}

\newcommand{\norm}[1]{\left\|#1\right\|}

\newcommand{\Vun}[1]{{\boldsymbol{1}}_{#1}}
\newcommand{\Vzero}{\boldsymbol{0}}
\newcommand{\Id}[1]{\textbf{I}_{#1}}
\newcommand{\Indicfun}[2]{\textbf{1}_{#1}\left(#2\right)}

\newenvironment{algogo}[1]{
\smallskip
\noindent \hrule\vspace{0.2\baselineskip} \hrule
\smallskip
\begin{small}
\refstepcounter{algo} \center{\bf \textsc{Algorithm \thealgo:}}
\\{\center{\bf #1}}
\smallskip
\flushleft
 } {
\end{small}
\bigskip
\hrule\vspace{0.2\baselineskip} \hrule
\smallskip
}

\newcounter{algo}
\renewcommand{\thealgo}{\arabic{algo}}

\title{Joint Bayesian endmember extraction and linear unmixing for hyperspectral imagery}

\author{Nicolas Dobigeon$^{1,2}$, Sa\"id Moussaoui$^{3}$, Martial Coulon$^{1}$, \\
    Jean-Yves Tourneret$^{1}$ and Alfred O. Hero$^{2}$\\
\normalsize $^1$ University of Toulouse, IRIT/INP-ENSEEIHT, 2 rue Camichel, 31071 Toulouse, France. \\
\normalsize $^2$ University of Michigan, Department of EECS, Ann Arbor, MI 48109-2122, USA \\
\normalsize $^3$ IRCCyN - CNRS UMR 6597, ECN, 1 rue de la No\"e, 44321 Nantes Cedex 3, France \\
\small\texttt{\{Nicolas.Dobigeon, Jean-Yves.Tourneret,
Martial.Coulon\}@enseeiht.fr, said.moussaoui@irccyn.ec-nantes.fr,
hero@umich.edu}
\thanks{Part of this work is funded by GdR-ISIS/CNRS,
    a DGA fellowship from French Ministry of Defence
    and AFOSR grant FA9550-06-1-0324.}}

\begin{document}

\maketitle

\hyphenation{hie-rar-chi-cal}

\begin{abstract}
This paper studies a fully Bayesian algorithm for endmember
extraction and abundance estimation for hyperspectral imagery. Each
pixel of the hyperspectral image is decomposed as a linear
combination of pure endmember spectra following the linear mixing
model. The estimation of the unknown endmember spectra is conducted
in a unified manner by generating the posterior distribution of
abundances and endmember parameters under a hierarchical Bayesian
model. This model assumes conjugate prior distributions for these
parameters, accounts for non-negativity and full-additivity
constraints, and exploits the fact that the endmember proportions
lie on a lower dimensional simplex. A Gibbs sampler is proposed to
overcome the complexity of evaluating the resulting posterior
distribution. This sampler generates samples distributed according
to the posterior distribution and estimates the unknown parameters
using these generated samples. The accuracy of the joint Bayesian
estimator is illustrated by simulations conducted on synthetic and
real AVIRIS images.

\end{abstract}



\section{Introduction}
Over the last several decades, much research has been devoted to the
spectral unmixing problem. Spectral unmixing is an efficient way to
solve standard problems encountered in hyperspectral imagery. These
problems include pixel classification \cite{Chang2003}, material
quantification \cite{Plaza2005} and subpixel detection
\cite{Manolakis2001}. Spectral unmixing consists of decomposing a
pixel spectrum into a collection of material spectra, usually
referred to as \emph{endmembers}, and estimating the corresponding
proportions or \emph{abundances} \cite{Keshava2002}. To describe the
mixture, the most frequently encountered model is the macroscopic
model which gives a good approximation in the reflective spectral
domain ranging from $0.4\mu$m to $2.5\mu$m \cite{Singer1979}. The
linearization of the non-linear intimate model proposed by Hapke in
\cite{Hapke1981} results in this macroscopic model
\cite{Johnson1983}. The macroscopic model assumes that the observed
pixel spectrum is a weighted linear combination of the endmember
spectra.

As reported in \cite{Keshava2002}, linear spectral mixture analysis
(LSMA) has often been handled as a two step procedure: the endmember
extraction step  and the inversion step, respectively. In the first
step of analysis, the macroscopic materials that are present in the
observed scene are identified by using an Endmember Extraction
Algorithm (EEA). The most popular EEAs include PPI
\cite{Boardman1993} and N-FINDR \cite{Winter1999}, that apply a
linear model for the observations with non-negativity and
full-additivity\footnote{The full-additivity constraint, that will
be detailed in the following section, refers to a unit
$\ell_1$-norm.} constraints. This model results in endmember spectra
located on the vertices of a lower dimensional simplex. PPI and
N-FINDR estimate this simplex by identifying the largest simplex
contained in the data. Another popular alternative, called Vertex
Component Analysis (VCA) has been proposed in \cite{Nascimento2005}.
A common assumption in VCA, PPI and N-FINDR is that they require
pure pixels to be present in the observed scene, where pure pixels
are pixels composed of a single endmember.
Alternatively, Craig has proposed the Minimum Volume Transform (MVT)
to find the smallest simplex that contains all the pixels
\cite{Craig1994}. However, MVT-based methods (e.g. ORASIS
\cite{Bowles1995}) are not fully automated techniques: they provide
results that strongly depend on i) the algorithm initialization, ii)
some \emph{ad hoc} parameters that have to be selected by the user.
More generally, these previous EEAs avoid the difficult problem of
direct parameter estimation on the simplex. The interested reader is
invited to consult \cite{Plaza2004} and \cite{Martinez2006} for a
recent performance comparison of some standard EEAs.

The second step in LSMA, called the \emph{inversion} step, consists
of estimating the proportions of the materials identified by EEA
\cite{Keshava2000}. The inversion step can use various strategies
such as least square estimation \cite{Heinz2001}, maximum likelihood
estimation \cite{Settle1996} and Bayesian estimation
\cite{Dobigeon2008}.

The central premise of this paper is to propose an algorithm that
estimates the endmember spectra and their respective abundances
jointly in a single step. This approach  casts LSMA as a blind
source separation (BSS) problem \cite{Comon1991}. In numerous
fields, independent component analysis (ICA) \cite{Lee1998} has been
a mainstay approach to solve BSS problems. In hyperspectral imagery,
ICA has also been envisaged \cite{Bayliss1998}. However, as
illustrated in \cite{Keshava2000} and \cite{Dobigeon_SPIE_ERS_2005},
ICA may perform poorly for LSMA due to the strong dependence between
the different abundances \cite{Nascimento2005b}. Inspired by ICA,
dependent component analysis has been introduced in
\cite{Nascimento2007igarss} to exploit this dependance. However,
this approach assumes that the hyperspectral observations are
noise-free. Alternatively, non-negative matrix factorization (NMF)
\cite{Paatero1994} can also be used to solve BSS problem under
non-negativity constraints. In \cite{Sajda2004}, a NMF algorithm
that consists of alternately updating the signature and abundance
matrices has been successfully applied to identify constituent in
chemical shift imaging. In this work, the additivity constraint has
not been taken into account. Basic simulations conducted on
synthetic images show that such MNF strategies lead to weak
estimation performances. In \cite{Berman2004}, an iterative
algorithm called ICE (iterated constrained endmembers) is proposed
to minimize a penalized criterion. As noted in
\cite{Nascimento2007igarss}, results provided by ICE strongly depend
on the choice of the algorithm parameters. More recently, in
\cite{Miao2007}, Miao \emph{et al.} have proposed another iterated
optimization scheme performing NMF with an additivity constraint on
the abundance coefficients. However, as this constraint has been
included in the objective function, it is not necessarily ensured.
In addition the performances of the algorithm in \cite{Miao2007}
decrease significantly when the noise level increases.

The joint Bayesian model uses a Gibbs sampling algorithm to
efficiently solve the constrained spectral unmixing problem without
requiring the presence of pure pixels in the hyperspectral image. In
addition, to our knowledge, this is the first time that
non-negativity constraints for endmember spectra as well as hard
additivity and non-negativity constraints for the abundances are
jointly considered in hyperspectral imagery.

In many works, Bayesian estimation approaches have been adopted to
solve BSS problems (see for example \cite{Rowe2002}) like LSMA. The
Bayesian formulation allows one to directly incorporate constraints
into the model such as sparsity \cite{Fevotte2006}, non-negativity
\cite{Moussaoui2006} and full additivity (sum-to-one constraint)
\cite{Dobigeon2007ssp}. In this paper, prior distributions are
proposed for the abundances and endmember spectra to enforce the
constraints inherent to the hyperspectral mixing model. These
constraints include non-negativity and full-additivity of the
abundance coefficients (as in \cite{Dobigeon2008}) and
non-negativity of the endmember spectra.
The proposed joint LSMA approach is able to solve the endmember
spectrum estimation problem directly on a lower dimensional space
within a Bayesian framework. We believe that this is one of the
principal factors leading to performance improvements that we show
on simulated and real data in Sections~\ref{sec:simulations_synth}
and \ref{sec:simulations_real}. By estimating the parameters on the
lower dimensional space we effectively reduce the number of degrees
of freedom of the parameters relative to other methods (e.g.
\cite{Moussaoui2006}), translating into lower estimator bias and
variance. The problem of hyperparameter selection in our Bayesian
model is circumvented by adopting the hierarchical Bayesian approach
of \cite{Dobigeon2008} that produces a parameter-independent
Bayesian posterior distribution for the endmember spectra and
abundances.
To overcome the complexity of the full posterior distribution, a
Gibbs sampling strategy is derived to approximate standard Bayesian
estimators, e.g. the minimum mean squared error (MMSE) estimator.
Moreover, as the full posterior distribution of all the unknown
parameters is available, confidence interval can be easily computed.
These measures allow one to quantify the accuracy of the different
estimates.

The paper is organized as follows. The observation model is
described in Section~\ref{sec:LMM}. The different quantities
necessary to the Bayesian formulation are enumerated in
Section~\ref{sec:Bayesian_model}. Section~\ref{sec:Gibbs} presents
the proposed Gibbs sampler for joint abundance and endmember
estimation. Simulation results obtained with synthetic and real
AVIRIS data are reported in Sections~\ref{sec:simulations_synth} and
\ref{sec:simulations_real} respectively.
Section~\ref{sec:conclusions} concludes the paper. An appendix
provides details on our parameterization of the simplex and
selecting relevant and tractable priors.

\section{Linear mixing model and problem statement}\label{sec:LMM}
Consider $\nbpix$ pixels of an hyperspectral image acquired in
$\nbband$ spectral bands. According to the linear mixing model
(LMM), described for instance in \cite{Keshava2002}, the
$\nbband$-spectrum
$\Vpix{\nopix}=[\pix{\nopix}{1},\ldots,\pix{\nopix}{\nbband}]\transp$
of the $\nopix$th pixel ($\nopix=1,\ldots,\nbpix$) is assumed to be
a linear combination of $\nbmat$ spectra $\Vmat{\nomat}$ corrupted
by an additive Gaussian noise
\begin{equation}
\label{eq:LMM}
 \Vpix{\nopix} = \sum_{\nomat=1}^\nbmat
\Vmat{\nomat} \abond{\nopix}{\nomat} +\Vnoise{\nopix},
\end{equation}
where $\Vmat{\nomat} =
[\mat{\nomat}{1},\ldots,\mat{\nomat}{\nbband}]\transp$ denotes the
spectrum of the $\nomat$th material, $\abond{\nopix}{\nomat}$ is the
fraction of the $\nomat$th material in the ${\nopix}$th observation,
$\nbmat$ is the number of materials, $\nbband$ is the number of
available spectral bands and $\nbpix$ is the number of observations
(pixels). Moreover, in \eqref{eq:LMM}, $\Vnoise{\nopix} =
[\noise{\nopix}{1},\ldots,\noise{\nopix}{\nbband}]\transp$ is an
additive noise sequence which is assumed to be an independent and
identically distributed (i.i.d.) zero-mean Gaussian sequence with
covariance matrix $\MATnoisevar=\noisevar\Id{\nbband}$, where
$\Id{\nbband}$ is the identity matrix of dimension $\nbband \times
\nbband$
\begin{equation}
\label{eq:noise_model}
  \Vnoise{\nopix} \sim
\mathcal{N}\left(\Vzero_{\nbband},\MATnoisevar\right).
\end{equation}
The proposed model in \eqref{eq:noise_model} does not account for
any possible correlation in the noise sequences but has been widely
adopted in the literature \cite{Harsanyi1994,Chang1998,Chang1998b}.
However, some simulation results reported in paragraph
\ref{subsec:complex_noise} will show that the proposed algorithm is
robust to the violation of the i.i.d. noise assumption. Note finally
that the model in \eqref{eq:LMM} can be easily modified (see
\cite{DobigeonTR2008b}) to handle more complicated noise models with
different variances in each spectral band as in \cite{Li2004}, or by
taking into account correlations between spectral bands as in
\cite{Dobigeon2008}.

Due to physical considerations, described in \cite{Manolakis2001},
\cite{Dobigeon2008} or \cite{Chang2006}, the fraction vectors
$\Vabond{\nopix} = [
\abond{\nopix}{1},\ldots,\abond{\nopix}{\nbmat}]\transp$ in
\eqref{eq:LMM} satisfy the following non-negativity and
full-additivity (or \emph{sum-to-one}) constraints:
\begin{equation}
\label{eq:constraints} \left\{
\begin{array}{l}
\abond{\nopix}{\nomat}\geq0, \ \forall \nomat =1,\ldots,\nbmat,\\
\sum_{\nomat=1}^\nbmat \abond{\nopix}{\nomat} =1.
\end{array}\right.
\end{equation}
In other words, the $p$ abundance vectors belong to the space
\begin{equation}
\label{eq:Aspace}
  \calA = \left\{\Vabond{}: \  \norm{\Vabond{}}_1    = 1 \ \text{and } \Vabond{} \succeq \Vzero
  \right\},
\end{equation}
where $\norm{\cdot}_1$ is the $\ell_1$ norm $\norm{\bfx}_1 = \sum_i
\left|x_i\right|$, and $\Vabond{} \succeq \Vzero$ stands for the set
of inequalities $\left\{a_\nomat \geq
0\right\}_{\nomat=1,\ldots,\nbmat}$. Moreover, the endmember spectra
component $\mat{\nomat}{\noband}$ must satisfy non-negativity
constraints:
\begin{equation}
  \label{eq:constraints_spectra}
   \mat{\nomat}{\noband}\geq0, \ \forall \nomat=1,\ldots,\nbmat, \ \forall \noband =1,\ldots,\nbband.
\end{equation}

\begin{figure}[h!]
  \centering
  \includegraphics[width=\figwidth]{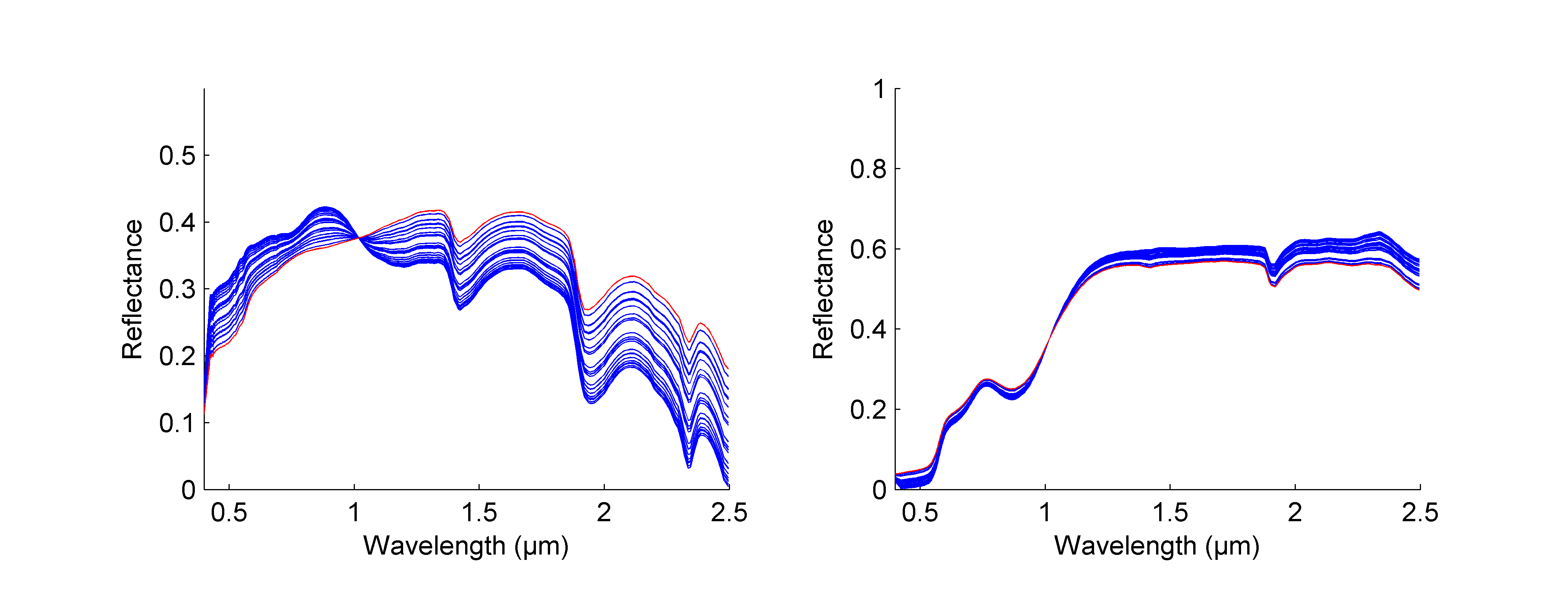}
  \caption{Range of admissible solution for two endmember spectra : construction concrete (left) and red brick (right).
  The actual endmember (red lines) are mixed according \eqref{eq:LMM} under the constraints in \eqref{eq:constraints} with random proportions to obtain
  $\nbpix=2500$ pixels. $50$ admissible solutions (blue lines) of the BSS
  problems in \eqref{eq:BSS_problem} are generated using
  \cite{Moussaoui2005icassp}.  }\label{fig:admis_sol}
\end{figure}

Considering all pixels, standard matrix notations yield:
\begin{equation}
\label{eq:BSS_problem}
  \MATpix = \MATmat\MATabond + \MATnoise,
\end{equation}
where
\begin{equation}
\begin{array}{llll}
  \MATpix   &= \left[\Vpix{1},\ldots,\Vpix{\nbpix}\right], &  \quad \MATmat &= \left[\Vmat{1},\ldots,\Vmat{\nbmat}\right], \\
  \MATabond &= \left[\Vabond{1},\ldots,\Vabond{\nbpix}\right], & \quad \MATnoise &=
  \left[\Vnoise{1},\ldots,\Vnoise{\nbpix}\right].
\end{array}
\end{equation}
In this work, we propose to estimate $\MATabond$ and $\MATmat$ from
the noisy observations $\MATpix$ under the constraints in
\eqref{eq:constraints} and \eqref{eq:constraints_spectra}. Note that
the unconstrained BSS problem for estimating M and A from Y is
ill-posed: if $\left\{\MATpix,\MATabond\right\}$ is an admissible
estimate then
$\left\{\MATpix\mathbf{H},\mathbf{H}\transp\MATabond\right\}$ is
also admissible for any unitary matrix H. In the LSMA problem, this
non-uniqueness can be partially circumvented by additional
constraints such as full-additivity, which enables one to handle the
scale indeterminacy. Consequently, these unit $\ell_1$-norm
constraints on the abundance vectors avoid using more complex
strategies for direct estimation of the scale \cite{Veit2009}.
Despite the constraints in \eqref{eq:constraints} and
\eqref{eq:constraints_spectra}, uniqueness of the couple
$\left\{\MATmat,\MATabond\right\}$ solution of the LSMA
\eqref{eq:BSS_problem} is not systematically ensured. To illustrate
this problem, $50$ admissible solutions\footnote{Admissible
solutions refer to couples $\left\{\MATmat,\MATabond\right\}$ that
satisfy \eqref{eq:constraints} and \eqref{eq:constraints_spectra}
and that follow the model \eqref{eq:LMM} in the noise-free case.}
have been depicted in Fig.~\ref{fig:admis_sol} for $\nbmat=2$
endmembers involved in the mixing of $\nbpix=2500$ pixels
\cite{Moussaoui2005icassp}. In the following section, the Bayesian
model used for the LSMA is presented.

\section{Bayesian model}\label{sec:Bayesian_model}
\subsection{Likelihood}
The linear mixing model defined in \eqref{eq:LMM} and the
statistical properties in \eqref{eq:noise_model} of the noise vector
$\Vnoise{\nopix}$ result in a conditionally Gaussian distribution
for the observation of the $p$th pixel: $\Vpix{\nopix}| \MATmat,
\Vabond{\nopix},\noisevar \sim \mathcal{N}\left(\MATmat
\Vabond{\nopix},\noisevar \Id{\nbband}\right)$. Therefore, the
likelihood function of $\Vpix{\nopix}$ can be expressed as
\begin{equation}
f\left(\Vpix{\nopix}\big|\MATmat, \Vabond{\nopix},\noisevar\right) =
\left(\frac{1}{2 \pi
\noisevar}\right)^{\frac{\nbband}{2}}\exp\left[-\frac{\left\|\Vpix{\nopix}-\MATmat\Vabond{\nopix}\right\|^2}{2\noisevar}\right],
\end{equation}
where $\left\|\bold{x} \right\|=\left({\bold{x}\transp
\bold{x}}\right)^{\frac{1}{2}}$ is the $\ell_2$ norm. Assuming the
independence between the noise sequences $\Vnoise{\nopix}$ ($\nopix
= 1,\ldots,\nbpix$), the likelihood function of all the observations
$\MATpix$ is:
\begin{equation}
\label{eq:likelihood}
  f\left(\MATpix \big| \MATmat, \MATabond, \noisevar \right)
  = \prod_{\nopix=1}^{\nbpix} f\left(\Vpix{\nopix}\big|\MATmat,
\Vabond{\nopix},\noisevar\right).
\end{equation}

\subsection{Prior model for the endmember spectra}
\subsubsection{Dimensionality reduction}
\label{subsubsec:dim_reduction} It is interesting to note that the
unobserved matrix $\bfX= \MATmat\MATabond = \MATpix - \MATnoise$ is
rank deficient under the linear model \eqref{eq:LMM}. More
precisely, the set
\begin{equation}
    \label{eq:polytope_S}
    \Simplex_{\MATmat} =\left\{\bfx \in \R^{\nbband} ;\  \bfx =
\sum_{\nomat=1}^{\nbmat} \lambda_{\nomat}\Vmat{\nomat},\
\sum_{\nomat=1}^{\nbmat} \lambda_{\nomat} = 1, \
 \lambda_{\nomat} \geq 0\right\}
\end{equation}
is a $(\nbmat-1)$-dimensional convex polytope of $\R^{\nbband}$
whose vertices are the $\nbmat$ endmember spectra $\Vmat{\nomat}$
($\nomat=1,\ldots,\nbmat$) to be recovered. Consequently, in the
noise-free case, $\bfX$ can be represented in a suitable
lower-dimensional subset $\calV_\nbpband$ of $\R^{\nbpband}$ without
lost of information.
\begin{figure}[h!]
  \centering
  \includegraphics[width=\figwidth]{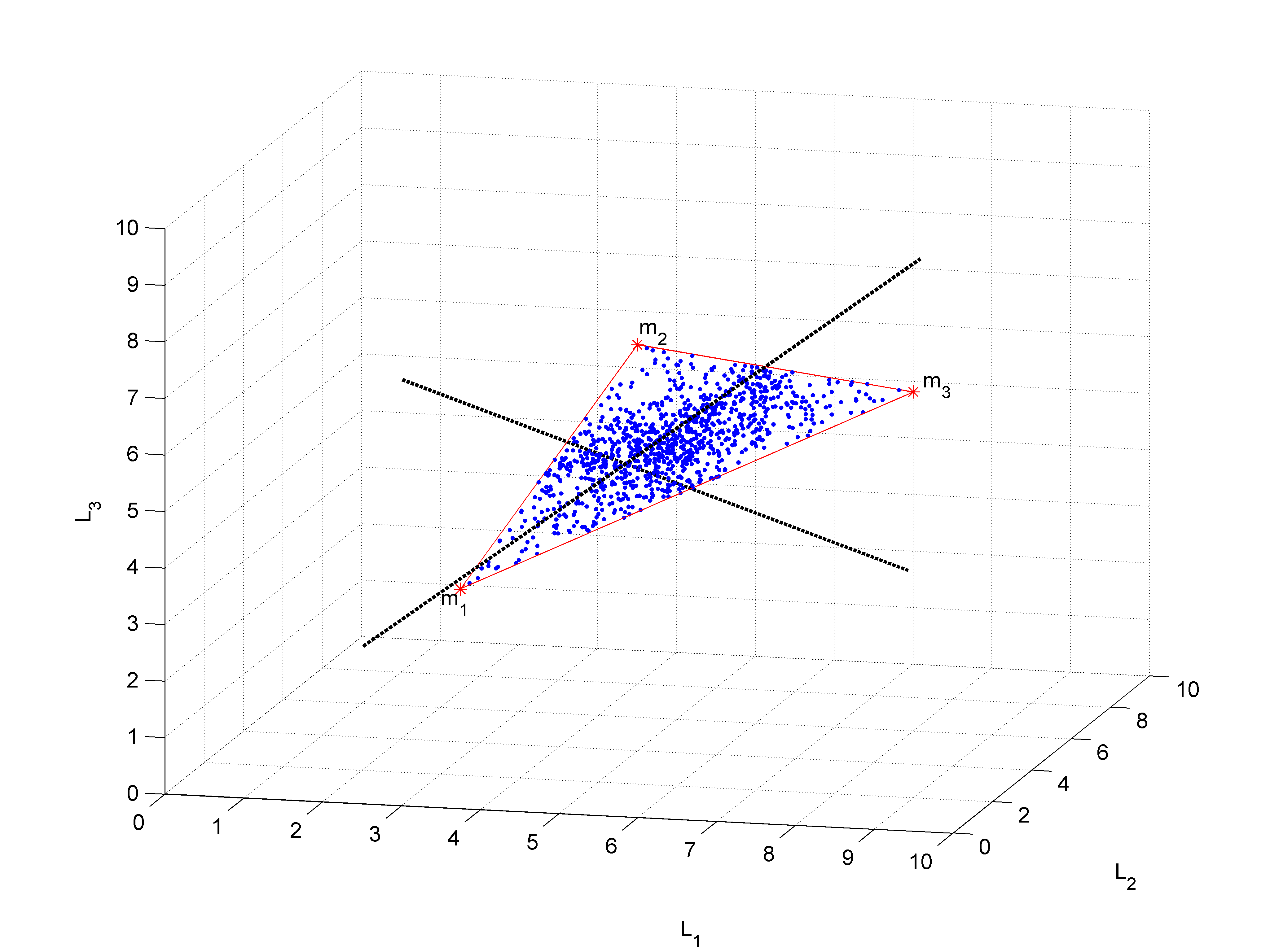}
  \caption{Example of hyperspectral data observed in $3$ spectral bands. The mixed pixels (blue points)
    belong to the $\nbmat$-dimensional convex polytope $\Simplex_{\MATmat}$ (red lines) whose vertices are the endmembers
    spectra $\Vmat{1},\ldots,\Vmat{\nbmat}$ (red stars).
  The first two principal axes estimated by a PCA appear in dashed lines and define the projection subset $\calV_{\nbpband}$.}    \label{fig:datasubspace}
\end{figure}
To illustrate this property, $\nbpix=1000$ pixels resulting from a
noise-free mixture of $\nbmat=3$ endmembers are represented in
Fig.~\ref{fig:datasubspace}. As noted in \cite{Keshava2002}, this
dimensionality reduction is a common step of the LSMA, adopted by
numerous EEAs, such as N-FINDR \cite{Winter1999} or PPI
\cite{Boardman1993}. Similarly, we propose to estimate the
projection $\Vpmat{\nomat}$ ($\nomat=1,\ldots,\nbmat$) of the
endmember spectra $\Vmat{\nomat}$ in the subspace $\calV_\nbpband$.
The identification of this subspace can be achieved via a standard
dimension reduction procedure. In the sequel, we propose to define
$\calV_\nbpband$ as the subspace spanned by $\nbpband$ orthogonal
axes $\bfv_1,\ldots,\bfv_{\nbpband}$ identified by a principal
component analysis (PCA) on the observations $\MATpix$
\cite{Jolliffe1986}:
\begin{equation}
 \label{eq:Vspace}
  \calV_\nbpband = \textrm{span}\left(\bfv_1,\ldots,\bfv_{\nbpband}\right).
\end{equation}
The first two principal axes are identified in
Fig.~\ref{fig:datasubspace} for the synthetic hyperspectral data. In
the following paragraph, the PCA is described. Note however that
this PCA-based dimension reduction step can be easily replaced by
other projection techniques, such as the maximum noise fraction
(MNF) transform \cite{Green1988} that has been considered in
paragraph \ref{subsec:complex_noise}.

\subsubsection{PCA projection}\label{subsubsec:PCA}
 The $\nbband \times \nbband $
empirical covariance matrix $\boldsymbol{\Upsilon}$ of the data
$\MATpix$ is given by:
\begin{equation}
\label{eq:empirical_covariance} \boldsymbol{\Upsilon} =
\frac{1}{\nbpix} \sum_{\nopix=1}^{\nbpix}
\left(\Vpix{\nopix}-\bar{\Vpix{}}\right)\left(\Vpix{\nopix}-\bar{\Vpix{}}\right)\transp
\end{equation}
where $\bar{\Vpix{}}$ is the empirical mean:
\begin{equation}
\label{eq:empirical_mean} \bar{\Vpix{}} =
\frac{1}{\nbpix}\sum_{\nopix=1}^{\nbpix}\Vpix{\nopix}.
\end{equation}
Let
\begin{equation}
\left\{
\begin{split}
    \label{eq:eigen_matrices}
  \bfD &=\mathrm{diag}\left(\lambda_1,\ldots,\lambda_{\nbpband}\right),\\
  \mathbf{V} &= \left[\bfv_1,\ldots,\bfv_{\nbpband}\right]\transp
\end{split}
\right.
\end{equation}
denote respectively the diagonal matrix of the $\nbpband$ highest
eigenvalues and the corresponding eigenvector matrix of
$\boldsymbol{\Upsilon}$. The PCA projection $\Vpmat{\nomat} \in
\R^{\nbpband}$ of the endmember spectrum $\Vmat{\nomat}\in
\R^{\nbband}$ is obtained as follows:
\begin{equation}
    \label{eq:tr_model}
  \Vpmat{\nomat} = \bfP \left(\Vmat{\nomat} -
  \bar{\Vpix{}}\right),
\end{equation}
with $\bfP = \mathbf{D}^{-\frac{1}{2}} \bfV$. Equivalently,
\begin{equation}
 \label{eq:tr_model_bis}
  \Vmat{\nomat} = \bfU \Vpmat{\nomat} +
  \bar{\Vpix{}},
\end{equation}
with $\bfU = \bfV^{\dag} \bfD^{\frac{1}{2}} $ where $\bfV^{\dag}
\triangleq \left(\bfV\transp \bfV\right)\inv \bfV\transp$ is the
pseudo-inverse of $\bfV$. Note that in the subspace
$\calV_{\nbmat-1}$ obtained for $\nbpband=\nbmat-1$, the vectors
$\left\{\Vpmat{\nomat}\right\}_{\nomat=1,\ldots,\nbmat}$ form a
simplex that standard EEAs such as N-FINDR \cite{Winter1999}, MVT
\cite{Craig1994} and ICE \cite{Berman2004} try to recover. In this
paper, we estimate the vertices $\Vpmat{\nomat}$
($\nomat=1,\ldots,\nbmat$) of this simplex using a Bayesian
approach. The Bayesian prior distributions for the projections
$\Vpmat{\nomat}$ ($\nomat=1,\ldots,\nbmat$) are introduced in the
following paragraph.

\subsubsection{Prior distribution for the projected spectra}
All the elements of the subspace $\calV_\nbpband$ may not be
appropriate projected spectra according to \eqref{eq:tr_model}.
Indeed, the $\nbpband \times 1$ vector $\Vpmat{\nomat}$ has to
ensure non-negativity constraints \eqref{eq:constraints_spectra} of
the corresponding reconstructed $\nbband \times 1$ spectra
$\Vmat{\nomat}$. For each endmember $\Vmat{\nomat}$, straightforward
computations establish that for any $\nomat=1,\ldots,\nbmat$
\begin{equation}
  \left\{\mat{\noband}{\nomat} \geq 0,\  \forall \noband = 1,\ldots,\nbband\right\} \quad \Leftrightarrow \quad \left\{\Vpmat{\nomat} \in
  \Tspace{\nomat}\right\},
\end{equation}
where the set $\Tspace{\nomat}\subset \calV_\nbpband$ is defined by
the following $\nbband$ inequalities
\begin{equation}
\label{eq:Tspace}
  \Tspace{\nomat}= \left\{\Vpmat{\nomat} ; \bar{y}_{\noband}+\sum_{\nopband=1}^{\nbpband} u_{\noband,\nopband}\pmat{\nopband}{\nomat}
  \geq 0, \noband=1,\ldots,\nbband\right\},
\end{equation}
with
$\bar{\Vpix{}}=\left[\bar{y}_{1},\ldots,\bar{y}_{\nbband}\right]\transp$
and $\bfU = \left[u_{\noband,\nopband}\right]$. A
conjugate\footnote{For the main motivations of choosing conjugate
priors, see for instance \cite[Chap. 3]{Robert2007}.} multivariate
Gaussian distribution (MGD)
$\calN_{\Tspace{\nomat}}\left(\bfe_{\nomat},\pmatvar{\nomat}\Id{\nbpband}\right)$
truncated on the set $\Tspace{\nomat}$ is chosen as prior
distribution for $\Vpmat{\nomat}$. The probability density function
(pdf) $\phi_{\Tspace{\nomat}}\left(\cdot\right)$ of this truncated
MGD is defined by:
\begin{equation}
\label{eq:pmat_prior_ind}
  \phi_{\Tspace{\nomat}}\left(\Vpmat{\nomat}\big| \bfe_{\nomat},\pmatvar{\nomat}\Id{\nbpband} \right)
  \propto
 \phi\left(\Vpmat{\nomat}\big| \bfe_{\nomat},\pmatvar{\nomat}\Id{\nbpband} \right)\Indicfun{\Tspace{\nomat}}{\Vpmat{\nomat}} ,
\end{equation}
where $\propto$ stands for ``proportional to",
$\phi\left(\cdot|\bfu,\bfW\right)$ is the pdf of the standard MGD
$\mathcal{N}\left(\bfu,\bfW\right)$ with mean vector $\bfu$ and
covariance matrix $\bfW$, and $\Indicfun{\Tspace{\nomat}}{\cdot}$ is
the indicator function on the set ${\Tspace{\nomat}}$:
\begin{equation}
  \Indicfun{\Tspace{\nomat}}{\bfx} = \left\{
                                  \begin{array}{ll}
                                    1, & \hbox{if $\bfx\in {\Tspace{\nomat}}$ } ; \\
                                    0, & \hbox{overwise.}
                                  \end{array}
                                \right.
\end{equation}
The normalizing constant
$K_{\Tspace{\nomat}}\left(\bfe_{\nomat},\pmatvar{\nomat}\Id{\nbpband}\right)$
in \eqref{eq:pmat_prior_ind} is defined as follows:
\begin{equation}
  K_{\Tspace{\nomat}}\left(\bfe_{\nomat},\pmatvar{\nomat}\Id{\nbpband}\right) = \int_{\Tspace{\nomat}}
  \phi\left(\bfx|\bfe_{\nomat},\pmatvar{\nomat}\Id{\nbpband}\right) d\bfx.
\end{equation}

This paper proposes to choose the mean vectors $\bfe_{\nomat}$
($\nomat=1,\ldots,\nbmat$) in \eqref{eq:pmat_prior_ind} as the
projected spectra of pure components previously identified by EEA,
e.g., N-FINDR. The variances $\pmatvar{\nomat}$
($\nomat=1,\ldots,\nbmat$) reflect the degree of confidence given to
this prior information. When no additional knowledge is available,
these variances are fixed to large values:
$\pmatvar{1}=\ldots=\pmatvar{\nbmat}=50$.

By assuming \emph{a priori} independence of the vectors
$\Vpmat{\nomat}$ ($\nomat=1,\ldots, \nbmat$), the prior distribution
for the projected endmember matrix $\MATpmat =
\left[\Vpmat{1},\ldots,\Vpmat{\nbmat}\right]$ is
\begin{equation}
  \label{eq:pmat_prior}
  f\left(\MATpmat\mid \bfE, \Vpmatvar\right) = \prod_{\nomat=1}^{\nbmat}
\phi_{\Tspace{\nomat}}\left(\Vpmat{\nomat}\big|
\bfe_{\nomat},\pmatvar{\nomat}\Id{\nbpband} \right),
\end{equation}
where $\bfE = \left[\bfe_{1},\ldots,\bfe_{\nbmat}\right]$ and
$\Vpmatvar = \left[\pmatvar{1},\ldots,\pmatvar{\nbmat}\right]$.

\subsection{Abundance prior} \label{subsubsec:abund_prior} For each
observed pixel $\nopix$, with the full additivity constraint
in~\eqref{eq:constraints}, the abundance vectors $\Vabond{\nopix}$
($\nopix=1,\ldots,\nbpix$) can be rewritten as
\begin{equation*}
\Vabond{\nopix} = \left[
  \begin{array}{c}
    \Vcoeff{\nopix} \\
    \abond{\nopix}{\nbmat} \\
  \end{array}
\right] \qquad \text{with } \qquad\Vcoeff{\nopix} = \left[
  \begin{array}{c}
    \abond{\nopix}{1} \\
    \vdots\\
    \abond{\nopix}{\nbmat-1} \\
  \end{array}
\right],
\end{equation*}
and
$\abond{\nopix}{\nbmat}=1-\sum_{\nomat=1}^{\nbmat-1}\abond{\nopix}{\nomat}$.
Following the model in \cite{Dobigeon2008}, the priors chosen for
$\Vcoeff{\nopix}$ ($\nopix=1,\ldots,\nbpix$) are uniform
distributions on the simplex $\Simplex$ defined by:
\begin{equation}
    \label{eq:Simplex}
  \Simplex =\left\{ \Vcoeff{\nopix}; \  \norm{\Vcoeff{\nopix}}_1 \leq 1 \
\text{and } \Vcoeff{\nopix} \succeq \Vzero \right\}.
\end{equation}
Choosing this prior distribution for $\Vcoeff{\nopix}$
(${\nopix}=1,\ldots,\nbpix$) is equivalent to electing a Dirichlet
distribution $\calD\left(1,\ldots,1\right)$, i.e a uniform
distribution on $\calA$ defined in \eqref{eq:Aspace}, as prior
distribution for the full abundance vector $\Vabond{\nopix}$
\cite[Appendix A]{Robert2007}. However, the proposed
reparametrization will prove to be well adapted to the Gibbs
sampling strategy introduced in Section~\ref{sec:Gibbs}.

Under the assumption of statistical independence between the
abundance vectors $\Vcoeff{\nopix}$ ($\nopix = 1,\ldots,\nbpix$),
the full prior distribution for partial abundance matrix $\MATcoeff
= \left[\Vcoeff{1},\ldots,\Vcoeff{\nbpix}\right]\transp$ can be
written
\begin{equation}
\label{eq:abond_prior}
  f\left(\MATcoeff\right) \propto \prod_{\nopix=1}^{\nbpix}
  \Indicfun{\Simplex}{\Vcoeff{\nopix}}.
\end{equation}

As noted in \cite{Dobigeon2008}, the uniform prior distribution
reflects the lack of \emph{a priori} knowledge about the abundance
vector. Moreover, for the BSS problem here, this imposes a strong
constraint on the size of the simplex to be recovered. As
demonstrated in the Appendix, among two \emph{a priori} equiprobable
solutions of the BSS problem, this uniform prior allows one to favor
\emph{a posteriori} the solution corresponding to the polytope in
the projection subset $\calV_{\nbpband}$ with smallest volume. This
property has also been exploited in \cite{Craig1994}.

\subsection{Noise variance prior} A conjugate prior
is chosen for $\noisevar$:
\begin{equation}
\label{eq:var_prior} \noisevar\left|{\nu},{\gamma}\right. \sim
\mathcal{IG}\left(\frac{\nu}{2},\frac{\gamma}{2}\right),
\end{equation}
where $\mathcal{IG}\left(\frac{\nu}{2},\frac{\gamma}{2}\right)$
denotes the inverse-gamma distribution with parameters
$\frac{\nu}{2}$ and $\frac{\gamma}{2}$. As in previous works
(\cite{Punskaya2002} and \cite{Dobigeon2007b}), the hyperparameter
$\nu$ will be fixed to $\nu=2$. On the other hand, $\gamma$ will be
a random and adjustable hyperparameter, whose prior distribution is
defined below.

\subsection{Prior distribution for hyperparameter $\gamma$}
The prior for $\gamma$ is a non-informative Jeffreys' prior
\cite{Jeffreys1961} which reflects the lack of knowledge regarding
this hyperparameter:
\begin{equation}
\label{eq:gamma_prior} f\left(\gamma\right) \propto
\frac{1}{\gamma}\Indicfun{\mathbb{R}^+}{\gamma}.
\end{equation}

\subsection{Posterior distribution}
The posterior distribution of the unknown parameter vector
$\paramvect=\left\{\MATcoeff,\MATpmat,\noisevar\right\}$ can be
computed from marginalization using the following hierarchical
structure
\begin{equation}
f( \paramvect | \MATpix ) = \int f( \paramvect, \gamma | \MATpix) d
\gamma \propto \int f(\MATpix|\paramvect) f(\paramvect | \gamma)
f(\gamma) d\gamma,
\end{equation}
where $f\left(\MATpix\big|\paramvect\right)$ and
$f\left(\gamma\right)$ are defined in \eqref{eq:likelihood} and
\eqref{eq:gamma_prior} respectively. Moreover, under the assumption
of \emph{a priori} independence between $\MATcoeff$, $\MATpmat$ and
$\noisevar$, the following result can be obtained:
\begin{equation}
f\left(\paramvect\big|\gamma\right) =
f\left(\MATcoeff\right)f\left(\MATpmat\mid\bfE,\Vpmatvar\right)f\left(\noisevar\mid\nu,\gamma\right),
\end{equation}
where $f\left(\MATcoeff\mid\bfE,\Vpmatvar\right)$,
$f\left(\MATpmat\right)$ and $f\left(\noisevar\mid\nu,\gamma\right)$
have been defined in Eq.'s~\eqref{eq:abond_prior},
\eqref{eq:pmat_prior} and \eqref{eq:var_prior}, respectively.

This hierarchical structure allows one to integrate out the
hyperparameter $\gamma$ from the joint distribution
$f\left(\paramvect,\gamma|\MATpix\right)$, yielding:
\begin{equation}
\begin{split}
\label{eq:posterior_full}
 f&\left(\MATcoeff,\MATpmat,\noisevar\big|\MATpix\right) \propto
    \prod_{\nopix=1}^\nbpix\Indicfun{\Simplex}{\Vcoeff{\nopix}}\\
    &\times \prod_{\nomat=1}^\nbmat\exp\left[-\frac{\norm{\Vpmat{\nomat}-\bfe_{\nomat}}^2}{2\pmatvar{\nomat}}\right]\Indicfun{\Tspace{\nomat}}{\Vpmat{\nomat}}\\
    &\times \prod_{\nopix=1}^\nbpix\left[\left(\frac{1}{\noisevar}\right)^{\frac{\nbband}{2}+1}
    \exp\left(-\frac{\norm{\Vpix{\nopix}-\left(\bfU \MATpmat+\bar{\Vpix{}}\right)\Vabond{\nopix}}^2}{2\noisevar}\right)\right].
\end{split}
\end{equation}
Deriving the Bayesian estimators (e.g., MMSE or MAP) from the
posterior distribution in \eqref{eq:posterior_full} remains
intractable. In such case, it is very common to use Markov chain
Monte Carlo (MCMC) methods to generate samples asymptotically
distributed according to the posterior distribution. The Bayesian
estimators can then be approximated using these samples. The next
section studies a Gibbs sampling strategy allowing one to generate
samples distributed according to \eqref{eq:posterior_full}.

\section{Gibbs sampler}\label{sec:Gibbs}
Random samples (denoted by ${\cdot}^{(t)}$ where $t$ is the
iteration index) can be drawn from
$f\left(\MATcoeff,\MATpmat,\noisevar\mid\MATpix\right)$ using a
Gibbs sampler \cite{Robert1999}. This MCMC technique consists of
generating samples
$\left\{\MATcoeff^{(t)},\MATpmat^{(t)},{\boldsymbol{\sigma}}^{2(t)}\right\}$
distributed according to the conditional posterior distributions of
each parameter.

\begin{figure}[h!]
\begin{algogo}{Gibbs sampling algorithm for LSMA}
    \label{algo:Gibbs}
    \begin{itemize}
    \item \underline{Preprocessing:}
        \begin{itemize}
            \item Compute the empirical mean vector $\bar{\Vpix{}}$ in \eqref{eq:empirical_mean},
            \item Compute the matrices $\bfD$ and $\bfV$ in \eqref{eq:eigen_matrices} via a PCA,
            \item Set $\bfU=\left(\bfV\transp \bfV\right)\inv \bfV\transp
            \bfD^{\frac{1}{2}}$,
            \item For $\nomat=1,\ldots,\nbmat$, choose the \emph{a priori} estimated endmembers $\bfe_{\nomat} \in \calV_{\nbpband}$ in \eqref{eq:pmat_prior_ind},
        \end{itemize}
    \item \underline{Initialization:}
        \begin{itemize}
            \item For $\nomat=1,\ldots,\nbmat$, sample $\Vpmat{\nomat}^{(0)}$ from \eqref{eq:pmat_prior_ind},
            \item For $\nomat=1,\ldots,\nbmat$, set $\Vmat{\nomat}^{(0)} = \bfU\Vpmat{\nomat}^{(0)} +
                \bar{\Vpix{}}$,
            \item Sample $\sigma^{2(0)}$ from
                \eqref{eq:var_prior},
            \item Set $t \leftarrow 1$,
        \end{itemize}
    \item \underline{Iterations:} for $t=1,2, \ldots, $ do
        \begin{itemize}
            \item[1.] For $\nopix=1,\ldots,\nbpix$, sample
                $\Vcoeff{\nopix}^{(t)}$ from \eqref{eq:gene_abond},
            \item[2.] For $\nomat=1,\ldots,\nbmat$, for $k=1,\ldots,\nbpband$,
                    sample $\pmat{k}{\nomat}^{(t)}$ from \eqref{eq:posterior_pmap_coord},
            \item[3.] For $\nomat=1,\ldots,\nbmat$, set $\Vmat{\nomat}^{(t)} = \bfU\Vpmat{\nomat}^{(0)} +
                \bar{\Vpix{}}$,
            \item[4.] Sample $\sigma^{2(t)}$ from
                \eqref{eq:gene_noisevar}.
            \item[5.] Set $t \leftarrow t+1$.
        \end{itemize}
    \end{itemize}
\end{algogo}
\end{figure}

\subsection{Sampling from $f\left(\MATcoeff|\MATpmat,\noisevar,\MATpix\right)$}
\label{subsec:gene_abond} Straightforward computations yield for
each observation:
\begin{multline}
f\left(\Vcoeff{\nopix}\left|\MATpmat,\noisevar,\Vpix{\nopix}\right.\right)\\
\propto
    \exp\left[-\frac{\left(\Vcoeff{\nopix}-\boldsymbol{\upsilon}_\nopix\right)\transp{\boldsymbol{\Sigma}}^{-1}_\nopix
        \left(\Vcoeff{\nopix}-\boldsymbol{\upsilon}_\nopix\right)}{2}\right]\Indicfun{\Simplex}{\Vcoeff{\nopix}},
        \label{eq:posteriorA}
\end{multline}
where:
\begin{equation}
\label{eq:param_normal} \left\{
\begin{split}
{\boldsymbol{\Sigma}}_\nopix &=
    \left[
    \left(\MATmat_{\text{-}\nbmat}-\Vmat{\nbmat}\Vun{\nbmat-1}\transp\right)\transp
    \MATnoisevar\inv
    \left(\MATmat_{\text{-}\nbmat}-\Vmat{\nbmat}\Vun{\nbmat-1}\transp\right)\right]^{-1},\\
{\boldsymbol{\upsilon}}_\nopix& =
    {\boldsymbol{\Sigma}}_\nopix
    \left[
    \left(\MATmat_{\text{-}\nbmat}-\Vmat{\nbmat}\Vun{\nbmat-1}\transp\right)\transp
    \MATnoisevar\inv
    \left(\Vpix{\nopix}-\Vmat{\nbmat}\right)\right],
\end{split}
\right.
\end{equation}
with $\MATnoisevar\inv=\frac{1}{\noisevar}\Id{\nbband}$,
$\Vun{\nbmat-1}=[1,\ldots,1]\transp \in \mathbb{R}^{\nbmat-1}$ and
where $\MATmat_{\text{-}\nbmat}$ denotes the matrix $\MATmat$ whose
$\nbmat$th column has been removed. As a consequence,
$\Vcoeff{\nopix}\big|\MATpmat,\noisevar,\Vpix{\nopix}$ is
distributed according to an MGD truncated on the simplex $\Simplex$
in \eqref{eq:Simplex}:
\begin{equation}
\label{eq:gene_abond}
\Vcoeff{\nopix}\left|\MATpmat,\noisevar,\Vpix{\nopix}\right. \sim
\mathcal{N}_{\Simplex}\left(\boldsymbol{\upsilon}_{\nopix},\boldsymbol{\Sigma}_{\nopix}\right).
\end{equation}
Note that samples can be drawn from an MGD truncated on a simplex
using efficient Monte Carlo simulation strategies such as described
in \cite{DobigeonTR2007b}.

\subsection{Sampling from $f\left(\MATpmat|\MATcoeff,\noisevar,\MATpix\right)$}
\label{subsec:gene_pmat}

Define  $\MATpmat_{\text{-}\nomat}$ as the matrix $\MATpmat$ whose
$\nomat$th column has been removed. Then the conditional posterior
distribution of $\Vpmat{\nomat}$ ($\nomat=1,\ldots,\nbmat$) is:
\begin{multline}
  f\left(\Vpmat{\nomat} | \MATpmat_{\text{-}\nomat}, \Vcoeff{\nomat}, \noisevar,\MATpix\right) \propto\\
\exp\left[-\frac{1}{2}\left(\Vpmat{\nomat}-\boldsymbol{\tau}_\nomat\right)\transp
\boldsymbol{\Lambda}_{\nomat}\inv\left(\Vpmat{\nomat}-\boldsymbol{\tau}_\nomat\right)\right]\Indicfun{\Tspace{\nomat}}{\Vpmat{\nomat}},
\end{multline}
with
\begin{equation}
\label{eq:param_normal} \left\{
\begin{split}
{\boldsymbol{\Lambda}}_\nomat &=
                \left[\sum_{\nopix=1}^{\nbpix}\abond{\nopix}{\nomat}^2 \bfU\transp\MATnoisevar\inv\bfU
                + \frac{1}{\pmatvar{\nomat}}\Id{\nbpband}\right]\inv ,\\
{\boldsymbol{\tau}}_\nomat& =
                {\boldsymbol{\Lambda}}_\nomat
                \left[\sum_{\nopix=1}^{\nbpix}\abond{\nopix}{\nomat}^2\bfU\transp\MATnoisevar\inv\boldsymbol{\epsilon}_{\nopix,\nomat}
                + \frac{1}{\pmatvar{\nomat}}\bfe_{\nomat}
                \right],
\end{split}
\right.
\end{equation}
and
\begin{equation}
  \boldsymbol{\epsilon}_{\nopix,\nomat}= \Vpix{\nopix}  -
\abond{\nopix}{\nomat}\bar{\Vpix{}} -
    \sum_{j\neq\nomat} \abond{\nopix}{j}\Vmat{j}.
\end{equation}
Note that $\Vmat{j}=\bfU\Vpmat{j}+\bar{\Vpix{}}$. As a consequence,
the posterior distribution of $\Vpmat{\nomat} $ is the following
truncated MGD:
\begin{equation}
  \Vpmat{\nomat} \mid \MATpmat_{\text{-}\nomat}, \Vcoeff{\nomat}, \noisevar,\MATpix \sim
\calN_{\Tspace{\nomat}}\left(\boldsymbol{\tau}_\nomat,\boldsymbol{\Lambda}_{\nomat}\right).
\end{equation}
Generating vectors distributed according to this distribution is a
difficult task, mainly due to the truncation on the subset
$\Tspace{\nomat}$. An alternative consists of generating each
component $\pmat{\nopband}{\nomat}$ of $\Vpmat{\nomat}$
conditionally upon the others
$\Vpmat{\text{-}\nopband,\nomat}=\left\{\pmat{j}{\nomat}\right\}_{j\neq
k}$. More precisely, by denoting $\calU_\nopband^+=\left\{l;
u_{l,\nopband}>0\right\}$, $\calU_\nopband^-=\left\{l;
u_{l,\nopband}<0\right\}$ and $\varepsilon_{l,k,\nomat} =
\bar{y}_{\noband}+\sum_{j\neq k} u_{l,j}\pmat{j}{\nomat}$, one can
write
\begin{equation}
\label{eq:posterior_pmap_coord}
  \pmat{\nopband}{\nomat} |\Vpmat{\text{-}\nopband,\nomat}, \MATpmat_{\text{-}\nomat}, \Vcoeff{\nomat}, \noisevar,\MATpix
  \sim
  \calN_{\left[\pmat{\nopband}{\nomat}^-,\pmat{\nopband}{\nomat}^+\right]}\left(w_{{\nopband},{\nomat}},z^2_{{\nopband},{\nomat}}\right),
\end{equation}
with
\begin{equation}
\left\{
  \begin{split}
    \pmat{\nopband}{\nomat}^- &= \max_{l\in\calU_\nopband^+}-\frac{\varepsilon_{\noband,\nopband,\nomat}}{u_{\noband,\nopband}},\\
    \pmat{\nopband}{\nomat}^+ &= \min_{l\in\calU_\nopband^-}-\frac{\varepsilon_{\noband,\nopband,\nomat}}{u_{\noband,\nopband}},\\
  \end{split}
\right.
\end{equation}
and where $w_{{\nopband},{\nomat}}$ and $z^2_{{\nopband},{\nomat}}$
are the conditional mean and variance respectively, derived from the
partitioned mean vector and covariance matrix \cite[p. 324]{Kay1988}
(see \cite{DobigeonTR2007b} for similar computations). Generating
samples distributed according to the two-sided truncated Gaussian
distribution in \eqref{eq:posterior_pmap_coord} can be easily
achieved with the algorithm described in \cite{Robert1995}.

\subsection{Sampling from $f\left(\noisevar|\MATcoeff,\MATpmat,\MATpix\right)$}
\label{subsec:gene_noisevar}

The conditional distribution of
$\noisevar|\MATcoeff,\MATpmat,\MATpix$ is the following inverse
Gamma distribution:
\begin{equation}
\label{eq:gene_noisevar} \noisevar|\MATcoeff,\MATpmat,\MATpix \sim
\mathcal{IG}\left(\frac{\nbpix
\nbband}{2},\frac{1}{2}\sum_{\nopix=1}^{\nbpix}
\norm{\Vpix{\nopix}-\MATmat\Vabond{\nopix}}^2\right).
\end{equation}
Simulating according to this inverse Gamma distribution can be
achieved using a Gamma variate generator (see \cite[Ch.
9]{Devroye1986} and \cite[Appendix A]{Robert2007}).\\

To summarize, the hyperparameters that have to be fixed at the
beginning of the algorithm are the following: $\nu=2$,
$s^2_1=\ldots=s^2_\nbmat=50$ and
$\left\{\bfe_{\nomat}\right\}_{\nomat=1,\ldots,\nbmat}$ are set to
projected spectra identified by a standard EEA (e.g., N-FINDR).


\begin{figure*}
  \centering
  \includegraphics[width=\figwidthbis]{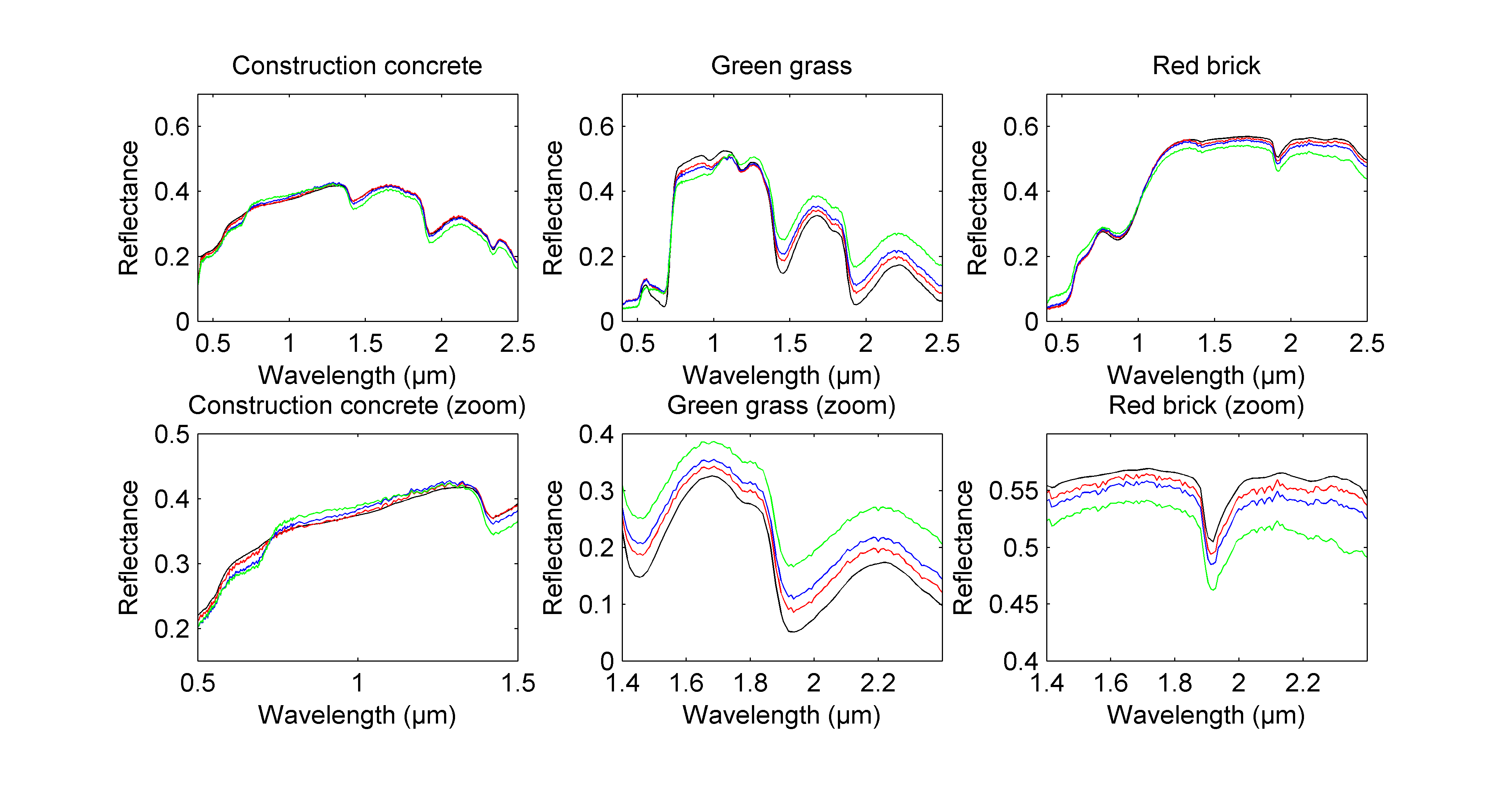}
  \caption{Actual endmembers (black lines),
    endmembers estimated by N-FINDR (blue lines), endmembers estimated by VCA (green lines) and
    endmembers estimated by proposed approach (red lines).}\label{fig:results_spectra}
\end{figure*}

\section{Simulations on synthetic data}\label{sec:simulations_synth}

To illustrate the accuracy of the proposed algorithm, simulations
are conducted on a $100 \times 100$ synthetic image. This
hyperspectral image is composed of three different regions with
$\nbmat=3$ pure materials representative of a sub-urban scene:
construction concrete, green grass and red brick. The spectra of
these endmembers have been extracted from the spectral libraries
distributed with the ENVI software \cite{ENVImanual2003} and are
represented in Fig.~\ref{fig:results_spectra} (top, black lines).
The reflectances are observed in $L=413$ spectral bands ranging from
$0.4\mu m$ to $2.5\mu m$. These $\nbmat=3$ components have been
mixed with proportions that have been randomly generated according
to truncated MGDs reflecting the means and variances reported in
Table~\ref{tab:image_parameters}. The generated abundance maps have
been depicted in Fig.~\ref{fig:abundance_maps} (top) in gray scale
where a white (resp. black) pixel stands for the presence (resp.
absence) of the material. The signal-to-noise ratio has been tuned
to $\textrm{SNR}_{\textrm{dB}}=15$dB.

\begin{table}
\renewcommand{\arraystretch}{1.2}
\begin{center}
\caption{Abundance means and variances of each endmember in each
region of the $100 \times 100$ hyperspectral image.}
\label{tab:image_parameters}
\begin{tabular}{|c|c|c|c|c|c|c|}
  \cline{2-7}
  \multicolumn{1}{c}{}&\multicolumn{2}{|c|}{Region $\#1$}  & \multicolumn{2}{|c|}{Region $\#2$} & \multicolumn{2}{|c|}{Region $\#3$}\\
  \cline{2-7}
  \multicolumn{1}{c|}{}&$\textrm{mean}$ & $\textrm{var.}$ & $\textrm{mean}$
                    & $\textrm{var.}$ & $\textrm{mean}$ & $\textrm{var.}$\\
  \hline
  {Endm. \#1} &   $0.60$ & $0.01$ & $0.25$ & $0.01$ & $0.25$ & $0.02$\\
  \hline
  {Endm. \#2}           &   $0.20$ & $0.02$ & $0.50$ & $0.01$ & $0.15$ & $0.005$\\
  \hline
  {Endm. \#3}             &   $0.20$ & $0.01$ & $0.25$ & $0.02$ & $0.60$ & $0.02$\\
  \hline
\end{tabular}
\end{center}
\end{table}

\subsection{Endmember spectrum estimation}
The resulting hyperspectral data have been unmixed by the proposed
algorithm. First, the space $\calV_\nbpband$ in \eqref{eq:Vspace}
has been identified by PCA as discussed in paragraph
\ref{subsubsec:PCA}. The hidden mean vectors $\bfe_{\nomat}$
($\nomat=1,\ldots,\nbmat$) of the normal distributions in
\eqref{eq:pmat_prior_ind} have been chosen as the PCA projections of
endmembers previously identified by N-FINDR. The hidden variances
$s^2_\nomat$ have all been chosen equal to
$s^2_1=\ldots=s^2_\nbmat=50$ to obtain vague priors (i.e. large
variances). The Gibbs sampler has been run with
$N_{\textrm{MC}}=1300$ iterations, including $N_{\textrm{bi}}=300$
burn-in iterations. The MMSE estimates of the abundance vectors
$\Vabond{\nopix}$ ($\nopix=1,\ldots,\nbpix$) and the projected
spectra $\Vpmat{\nomat}$ ($\nomat=1,\ldots,\nbmat$) have been
approximated by computing empirical averages over the last computed
outputs of the sampler
$\left\{\sampleVabond{\nomat}{t}\right\}_{t=1,\ldots,N_{\textrm{MC}}}$
and
$\left\{\sampleVpmat{\nomat}{t}\right\}_{t=1,\ldots,N_{\textrm{MC}}}$,
following the MMSE principle:
\begin{equation}
\label{eq:MMSE}
\begin{split}
  \hat{\bfx}_{\textrm{MMSE}} &= \mathrm{E}\left[ \bfx| \bfy\right]\\
  &\approx \frac{1}{N_{\textrm{MC}}-N_{\textrm{bi}}}
  \sum_{t=N_{\textrm{bi}}+1}^{N_{\textrm{MC}}} \bfx^{(t)}.
\end{split}
\end{equation}

\begin{figure}[h!]
  \centering
  \includegraphics[width=\figwidth]{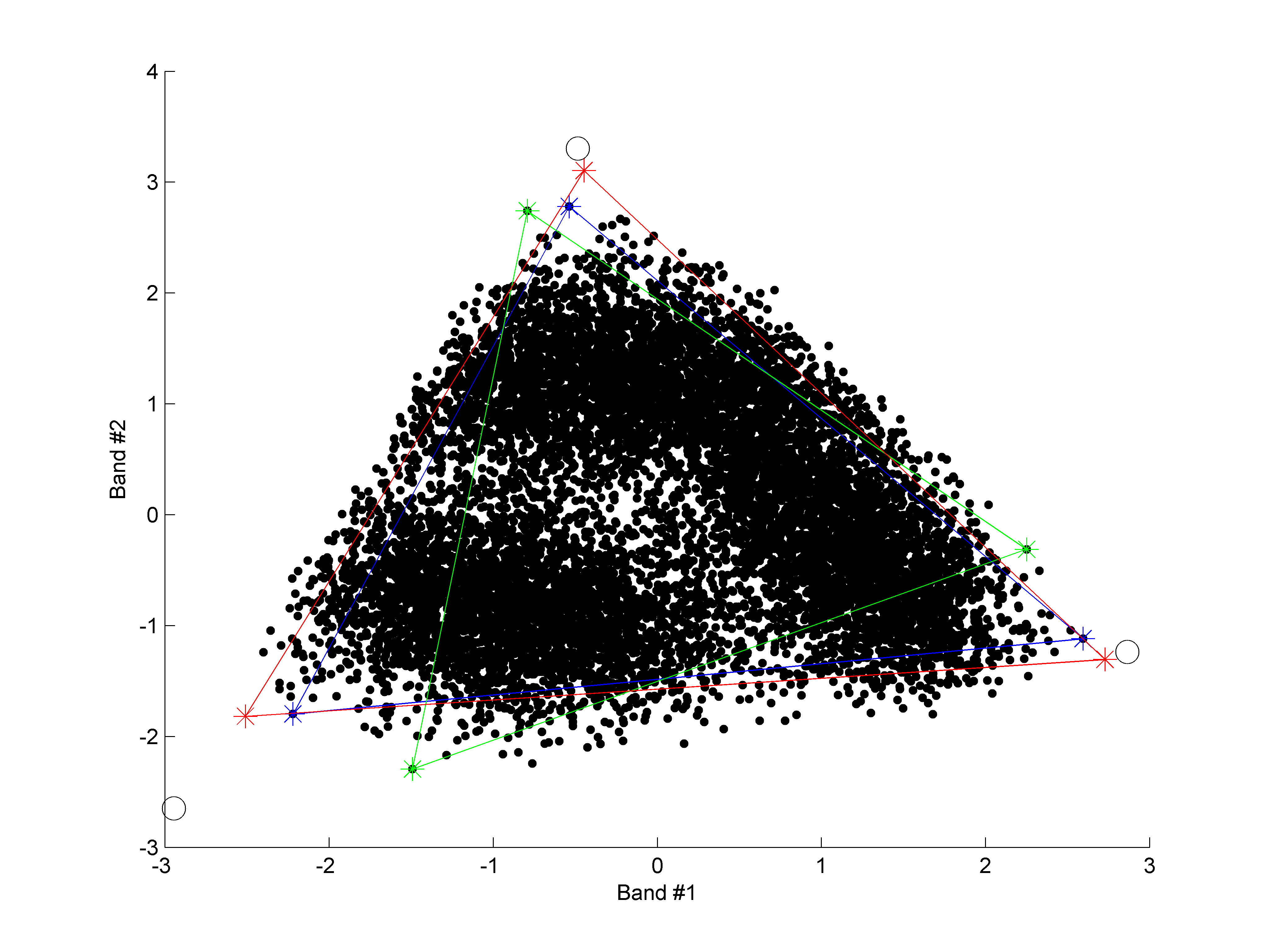}
  \caption{Scatter plot in the lower-dimensional space $\calV_2$: projected dataset (black points), actual endmembers (black circles),
    endmembers estimated by N-FINDR (blue stars), endmembers estimated by VCA (green stars) and endmembers estimated by proposed approach (red stars).
    All pixel spectra do not lie inside ground truth simplex due to simulated measurement noise.}\label{fig:scatter}
\end{figure}

The corresponding endmember spectra estimated by the proposed
algorithm are depicted in Fig.~\ref{fig:results_spectra} (top, red
lines). The proposed algorithm clearly outperforms N-FINDR and VCA,
as shown in Fig.~\ref{fig:results_spectra}. The scatter plot in
Fig.~\ref{fig:scatter} provides additional insight. The N-FINDR and
VCA algorithms assume the presence of pure pixels in the data.
However, as none of these pixels are pure, N-FINDR and VCA provide
poorer results than the proposed joint Bayesian algorithm. To
illustrate this point, the performances of the different algorithms
have been compared via two criteria. First, the mean square errors
(MSEs)
\begin{equation}
  \textrm{MSE}_r^2 =
  \left\|\hat{\Vmat{\nomat}}-{\Vmat{\nomat}}\right\|^2 ,\quad \nomat=1,\ldots,\nbmat
\end{equation}
are good quality indicators for the estimates. In addition, another
metric frequently encountered in hyperspectral imagery literature,
known as the spectral angle distance (SAD), has been considered. The
SAD measures the angle between the actual and the corresponding
estimated spectrum:
\begin{equation}
  \textrm{SAD}_r = \arccos\left(\frac{\langle \hat{\Vmat{\nomat}}, {\Vmat{\nomat}} \rangle}
        { \left\|\hat{\Vmat{\nomat}}\right\|
        \left\|{\Vmat{\nomat}}\right\|}\right),
\end{equation}
where $\langle \cdot,\cdot \rangle$ stands for the scalar product.
These performance criteria computed for the endmember spectra
estimated by the different algorithm have been reported in
Table~\ref{tab:performance_spectra}. They  show that the proposed
method performs significantly better than the others. The
computation times required by each of these algorithms are reported
in Table~\ref{tab:times} for a unoptimized MATLAB 2007b 32bit
implementation on a $2.2$GHz Intel Core 2. Obviously, the complexity
of the VCA and N-FINDR methods are lower than the proposed approach.
Note however that, contrary to the joint Bayesian procedure, these
standard EEA have to be coupled with an abundance estimation
algorithm. Moreover they only provide point estimates of the
endmember spectra. Note finally that the computational complexity of
N-FINDR, because it is combinatorial, increases drastically with the
number of pixels and endmembers.

\begin{table}[h!]
\renewcommand{\arraystretch}{1.35}
\begin{center}
\caption{Computational times of VCA, N-FINDR and the proposed
Bayesian method for unmixing $\nbpix = 32\times32$ pixels.}
\label{tab:times}
\begin{tabular}{|c|c|c|c|}
  \cline{2-4}
     \multicolumn{1}{c|}{} & {Bayesian}& {VCA}& {N-FINDR} \\
  \hline
   Times (s) &  $3250$ & $1$ & $23$\\
  \hline
\end{tabular}
\end{center}
\end{table}

\begin{table}
\renewcommand{\arraystretch}{1.35}
\begin{center}
\caption{Performance comparison between VCA, N-FINDR and the
proposed Bayesian method: $\textrm{MSE}^2$ and SAD $(\times
10^{-1})$ between the actual and the estimated endmember spectra.}
\label{tab:performance_spectra}
\begin{tabular}{|c|c|c|c|c|c|c|c|c|}
  \cline{3-9}
    \multicolumn{2}{c|}{} & \multirow{2}{*}{End.}  & \multicolumn{2}{|c|}{Bayesian} & \multicolumn{2}{|c|}{VCA} & \multicolumn{2}{|c|}{N-FINDR}\\
  \cline{4-9}
    \multicolumn{2}{c|}{} & \multirow{2}{*}{} & $\textrm{MSE}^2 $ & $\textrm{SAM}  $ & $\textrm{MSE}^2 $
                    & $\textrm{SAM} $ & $\textrm{MSE}^2 $ & $\textrm{SAM}$\\
  \hline
  \multirow{8}{*}{\rotatebox{90}{$\textrm{SNR}=5{\textrm{dB}}$}} & \multirow{3}{*}{\rotatebox{90}{$R=3$}}
                & \#1 &  $1.70$ & $0.63$ & $17.80$ & $1.91$ & $2.69$ & $0.75$\\
  & & \#2 &  $6.56$ & $1.49$ & $10.87$ & $1.80$ & $10.87$ & $1.80$\\
  & & \#3 &  $2.70$ & $0.59$ & $12.71$ & $1.40$ & $2.94$ & $0.60$\\
  \cline{2-9} & \multirow{5}{*}{\rotatebox{90}{$R=5$}}
     & \#1 &  $0.70$ & $0.01$ & $64.47$ & $3.93$ & $49.16$ & $3.46$\\
   & & \#2 &  $1.05$ & $0.49$ & $47.21$ & $3.13$ & $15.46$ & $2.07$\\
   & & \#3 &  $1.04$ & $0.49$ & $17.11$ & $1.81$ & $17.11$ & $1.81$\\
   & & \#4 &  $1.05$ & $0.49$ & $13.43$ & $1.23$ & $6.92$ & $0.79$\\
   & & \#5 &  $1.04$ & $0.49$ & $16.12$ & $1.45$ & $4.92$ & $0.88$\\
  \hline
  \hline
  \multirow{8}{*}{\rotatebox{90}{$\textrm{SNR}=15{\textrm{dB}}$}} & \multirow{3}{*}{\rotatebox{90}{$R=3$}}
     & \#1 &  $0.10$ & $0.15$ & $1.29$ & $0.48$ & $0.54$ & $0.33$\\
   & & \#2 &  $2.68$ & $0.92$ & $15.59$ & $2.12$ & $5.19$ & $1.26$\\
   & & \#3 &  $0.16$ & $0.12$ & $4.35$ & $0.71$ & $0.57$ & $0.22$\\
  \cline{2-9} & \multirow{5}{*}{\rotatebox{90}{$R=5$}}
    & \#1 &  $0.12$ & $0.17$ & $0.70$ & $0.40$ & $0.70$ & $0.40$\\
  & & \#2 &  $0.97$ & $0.52$ & $11.34$ & $1.44$ & $10.57$ & $1.68$\\
  & & \#3 &  $0.26$ & $0.22$ & $4.07$ & $0.72$ & $0.43$ & $0.25$\\
  & & \#4 &  $0.40$ & $0.15$ & $2.36$ & $0.44$ & $7.67$ & $0.47$\\
  & & \#5 &  $0.24$ & $0.18$ & $1.54$ & $0.43$ & $2.93$ & $0.60$\\
  \hline
  \hline
  \multirow{8}{*}{\rotatebox{90}{$\textrm{SNR}=25{\textrm{dB}}$}} & \multirow{3}{*}{\rotatebox{90}{$R=3$}}
     & \#1 &  $0.05$ & $0.09$ & $1.14$ & $0.52$ & $1.14$ & $0.52$\\
   & & \#2 &  $2.19$ & $0.83$ & $5.65$ & $1.33$ & $5.65$ & $1.33$\\
   & & \#3 &  $0.17$ & $0.14$ & $0.66$ & $0.22$ & $0.66$ & $0.22$\\
  \cline{2-9} & \multirow{5}{*}{\rotatebox{90}{$R=5$}}
    & \#1 &  $0.42$ & $0.29$ & $7.62$ & $1.32$ & $7.62$ & $1.32$\\
  & & \#2 &  $0.37$ & $0.34$ & $27.16$ & $2.23$ & $20.66$ & $2.40$\\
  & & \#3 &  $0.46$ & $0.29$ & $6.75$ & $1.10$ & $2.26$ & $0.65$\\
  & & \#4 &  $0.07$ & $0.09$ & $10.02$ & $0.93$ & $10.88$ & $0.70$\\
  & & \#5 &  $0.35$ & $0.20$ & $3.73$ & $0.61$ & $3.71$ & $0.62$\\
  \hline
\end{tabular}
\end{center}
\end{table}

\subsection{Abundance estimation}
\begin{figure}[h!]
  \centering
  \includegraphics[width=\figwidth]{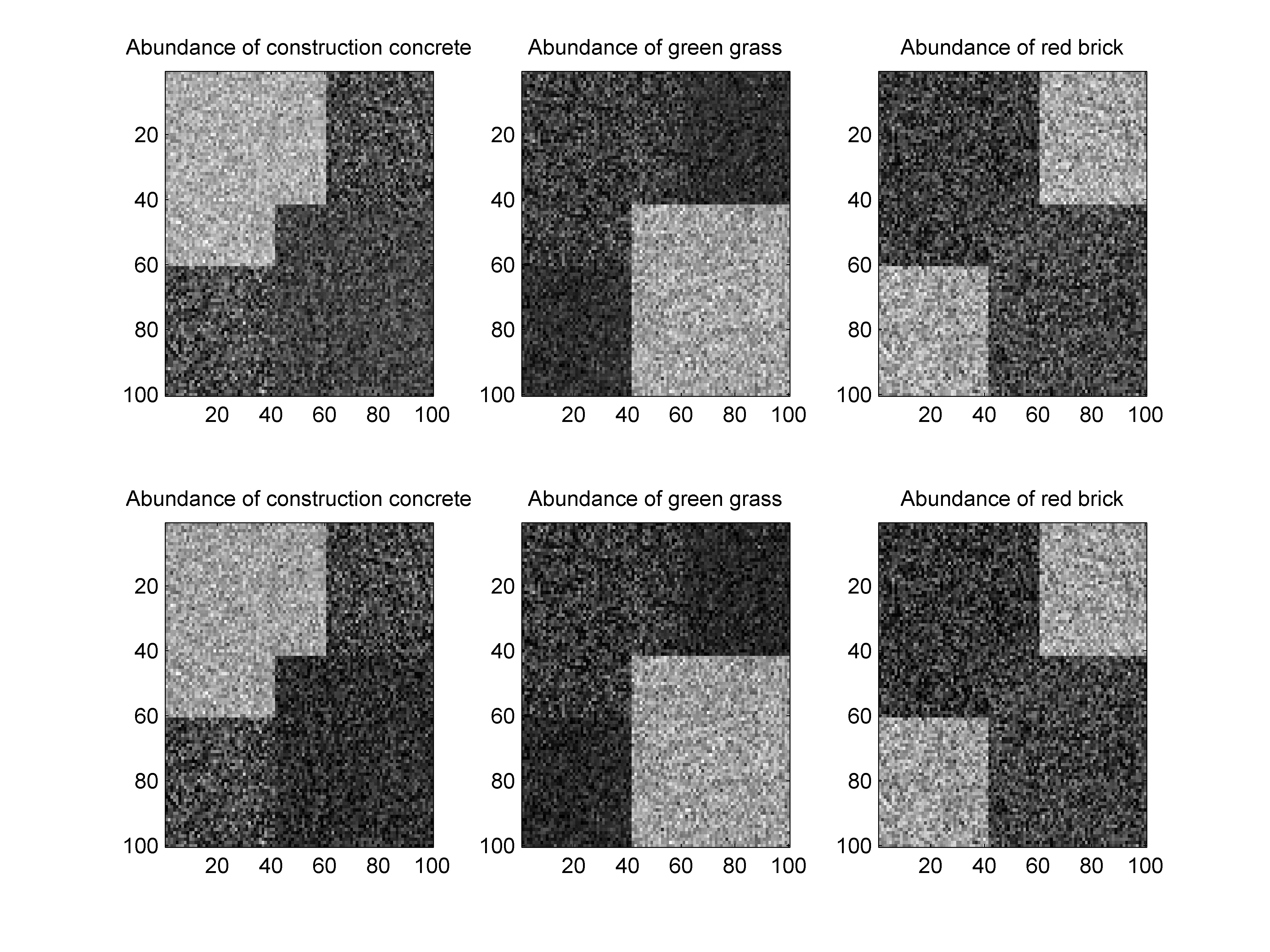}
  \caption{Top: actual endmember abundance maps. Bottom: estimated endmember abundance maps.}\label{fig:abundance_maps}
\end{figure}

The MMSE estimates of the abundance vectors for the $\nbpix=10^4$
pixels of the image have been computed following the MMSE principle
in \eqref{eq:MMSE}
\begin{equation}
  \hat{\Vabond{}}_{\nopix} =
  \frac{1}{N_{\textrm{MC}}-N_{\textrm{bi}}}
  \sum_{t=N_{\textrm{bi}}+1}^{N_{\textrm{MC}}} \sampleVabond{\nopix}{t}.
\end{equation}
The corresponding estimated abundance maps are depicted in
Fig.~\ref{fig:abundance_maps} (bottom) and are clearly in good
agreement with the simulated maps (top).

\begin{figure}[h!]
  \centering
  \includegraphics[width=\figwidth]{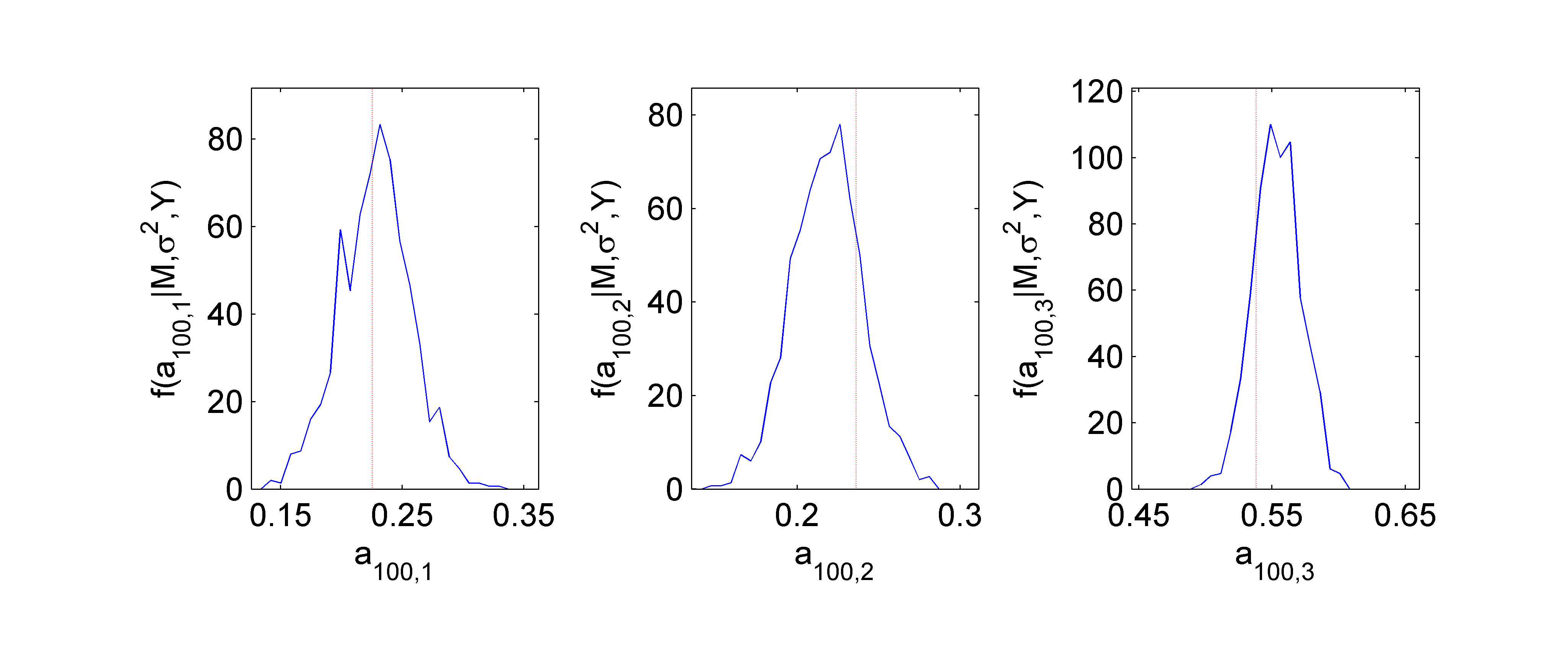}
  \caption{Posterior distributions of $\abond{\nopix}{r}$ ($r=1,\ldots,3$). The actual values are depicted in red
dotted lines.}\label{fig:posterior_abundance}
\end{figure}

Note that the proposed Bayesian estimation provides the joint
posterior distribution of the unknown parameters. Specifically,
these posteriors allow one to derive confidence intervals regarding
the parameters of interest. For instance, the posterior
distributions of the abundance coefficients is depicted in
Fig.~\ref{fig:posterior_abundance} for the pixel \#100. Note that
these estimated posteriors are in good agreement with the actual
values of $\Vabond{100}$ depicted in red dotted lines.

These results have been compared with estimates provided by the
N-FINDR or VCA algorithms, coupled with an abundance estimation
procedure based on the Fully Constrained Least-Squares (FCLS)
approach proposed by Heinz \emph{et al.} \cite{Heinz2001}. The
global abundance MSEs have been computed following:
\begin{equation}
  \textrm{GMSE}_{\nomat}^2 = \sum_{\nopix=1}^{\nbpix} \left(\hat{a}_{\nopix,\nomat}-\abond{\nopix}{\nomat}\right)^2,
\end{equation}
where $\hat{a}_{\nopix,\nomat}$ is the estimated abundance
coefficient of the material \#$\nomat$ in the pixel \#$\nopix$.
These performance measures have been reported in
Table~\ref{tab:performance_abundance} and confirm the accuracy of
the proposed Bayesian estimation method. Moreover, note that neither
N-FINDR nor VCA are able to provide confidence measures as those
depicted in Fig~\ref{fig:posterior_abundance}.

\begin{table}[h!]
\renewcommand{\arraystretch}{1.35}
\begin{center}
\caption{Performance comparison between VCA, N-FINDR and the
proposed Bayesian method: $\textrm{GMSE}^2$ between the actual and
the estimated abundance maps.} \label{tab:performance_abundance}
\begin{tabular}{|c|c|c|c|}
  \cline{2-4}
     \multicolumn{1}{c|}{} & {Bayesian}& {VCA}& {N-FINDR} \\
  \hline
   Endm. \#1 &  $25.68$ & $57.43$ & $30.66$\\
   Endm. \#2 &  $29.97$ & $74.48$ & $46.45$\\
   Endm. \#3 &  $3.19$ & $83.02$ & $11.22$\\
  \hline
\end{tabular}
\end{center}
\end{table}

\subsection{Other simulation scenarios}

Simulations have also been conducted with different noise levels
($\textrm{SNR}_{\textrm{dB}}=5\textrm{dB},~25\textrm{dB}$) and when
$\nbmat=3$ or $\nbmat=5$ endmembers are involved in the mixture.
Estimation performances for the VCA and N-FINDR algorithms, as well
as the proposed approach, have been summarized in
Table~\ref{tab:performance_spectra}. These results expressed in
terms of MSE and SAD corroborate the effectiveness of our Bayesian
estimation procedure.

\subsection{Robustness to non-i.i.d noise models}
\label{subsec:complex_noise} In this paragraph, we illustrate the
robustness of the proposed algorithm with respect to violation of
the i.i.d. noise assumption. More precisely, a so-called Gaussian
shaped noise inspired by \cite{Bioucas2008} has been considered. The
noise correlation matrix $\MATnoisevar =
\textrm{diag}\left(\sigma^2_1,\ldots,\sigma^2_\nbband\right)$ is
designed such that its diagonal elements $\sigma^2_\noband$
($\noband=1,\ldots,\nbband$) follow a Gaussian shape centered at
band $\nbband/2$
\begin{equation}
  \sigma^2_\noband = \noisevar
  \exp\left[-\frac{\left(\noband-L/2\right)^2}{2\eta^2}\right].
\end{equation}
The parameter $\noisevar$ can be tuned to choose the SNR whereas the
parameter $\eta$ adjusts the shape width ($\eta \rightarrow \infty$
corresponds to i.i.d. noise). For this simulation, the parameters
$\noisevar$ and $\eta$ have been fixed to $1.0\times10^{2}$ and $50$
respectively, leading to a noise level of
$\textrm{SNR}_{\textrm{dB}} = 15\textrm{dB}$.
%

When the noise is not i.i.d., dimensionality reduction methods based
on eigen-decomposition of observed data correlation matrix
$\boldsymbol{\Upsilon}$ introduced in
\eqref{eq:empirical_covariance} can be inefficient. In this case,
other hyperspectral subspace identification methods have to be
considered. Here the PCA-based dimension reduction step introduced
in paragraph~\ref{subsubsec:PCA} was replaced by two techniques: the
well-known MNF transform \cite{Green1988} approach and the more
recently introduced HySime algorithm \cite{Bioucas2008}. Both of
them requires noise covariance matrix $\MATnoisevar$ estimation,
which was implemented following \cite{Bioucas2008}. The estimation
performances for the proposed Bayesian estimation procedure coupled
with MNF or HySime are reported in
Table~\ref{tab:performance_spectra_colored} and compared with VCA
and N-FINDR. These results show that the proposed method i) can be
easily used with other dimension reduction procedures, and ii) is
quite robust to the i.i.d. noise assumption.

\begin{table*}
\renewcommand{\arraystretch}{1.35}
\begin{center}
\caption{Performance comparison between VCA, N-FINDR and the
proposed Bayesian method in presence of Gaussian shaped noise:
$\textrm{MSE}^2$ and SAD $(\times 10^{-1})$ between the actual and
the estimated endmember spectra.}
\label{tab:performance_spectra_colored}
\begin{tabular}{|c|c|c|c|c|c|c|c|c|}
  \cline{2-9}
     \multicolumn{1}{c}{} & \multicolumn{2}{|c|}{MNF + Bayesian} & \multicolumn{2}{|c|}{HySime + Bayesian}& \multicolumn{2}{|c|}{VCA}& \multicolumn{2}{|c|}{N-FINDR} \\
  \cline{2-9}
     \multicolumn{1}{c|}{} & $\textrm{MSE}^2$ & $\textrm{SAD} $ & $\textrm{MSE}^2$ & $\textrm{SAD} $ & $\textrm{MSE}^2 $
                    & $\textrm{SAD} $ & $\textrm{MSE}^2$ & $\textrm{SAD}$\\
  \hline
   Endm. \#1 &  $0.26$ & $0.25$ &  $0.42$ & $0.31$ & $1.11$ & $0.46$ & $1.11$ & $0.46$\\
   Endm. \#2 &  $1.99$ & $0.79$ &  $4.35$ & $1.16$ & $5.78$ & $1.33$ & $5.78$ & $1.33$\\
   Endm. \#3 &  $0.33$ & $0.19$ &  $0.57$ & $0.22$ & $1.94$ & $0.41$ & $2.19$ & $0.43$\\
  \hline
\end{tabular}
\end{center}
\end{table*}

\section{Real data}\label{sec:simulations_real}
This section illustrates the proposed algorithm on real
hyperspectral data. The considered hyperspectral image was acquired
over Moffett Field (CA, USA) in 1997 by the JPL spectro-imager
AVIRIS \cite{JPL_AVIRIS}. This image has been used in many works to
illustrate hyperspectral signal processing algorithms
\cite{Christophe2005,Akgun2005}.

\begin{figure}[h!]
  \centering
  \includegraphics[width=\figwidth]{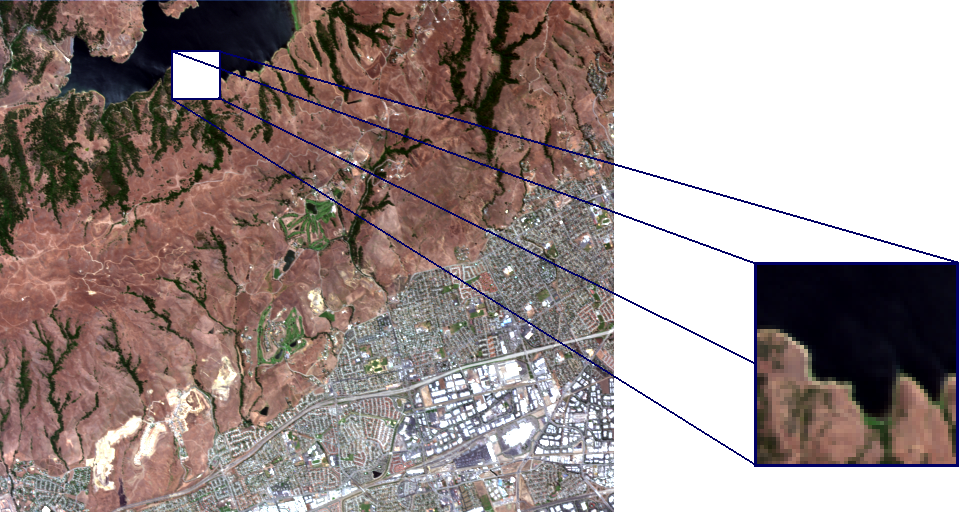}
  \caption{Real hyperspectral data: Moffett Field image acquired by AVIRIS in 1997 (left) and the region of interest (right)
  represented in synthetic colors.}\label{fig:Moffett_ROI}
\end{figure}

\begin{figure*}
  \centering
 \includegraphics[width=\figwidthbis]{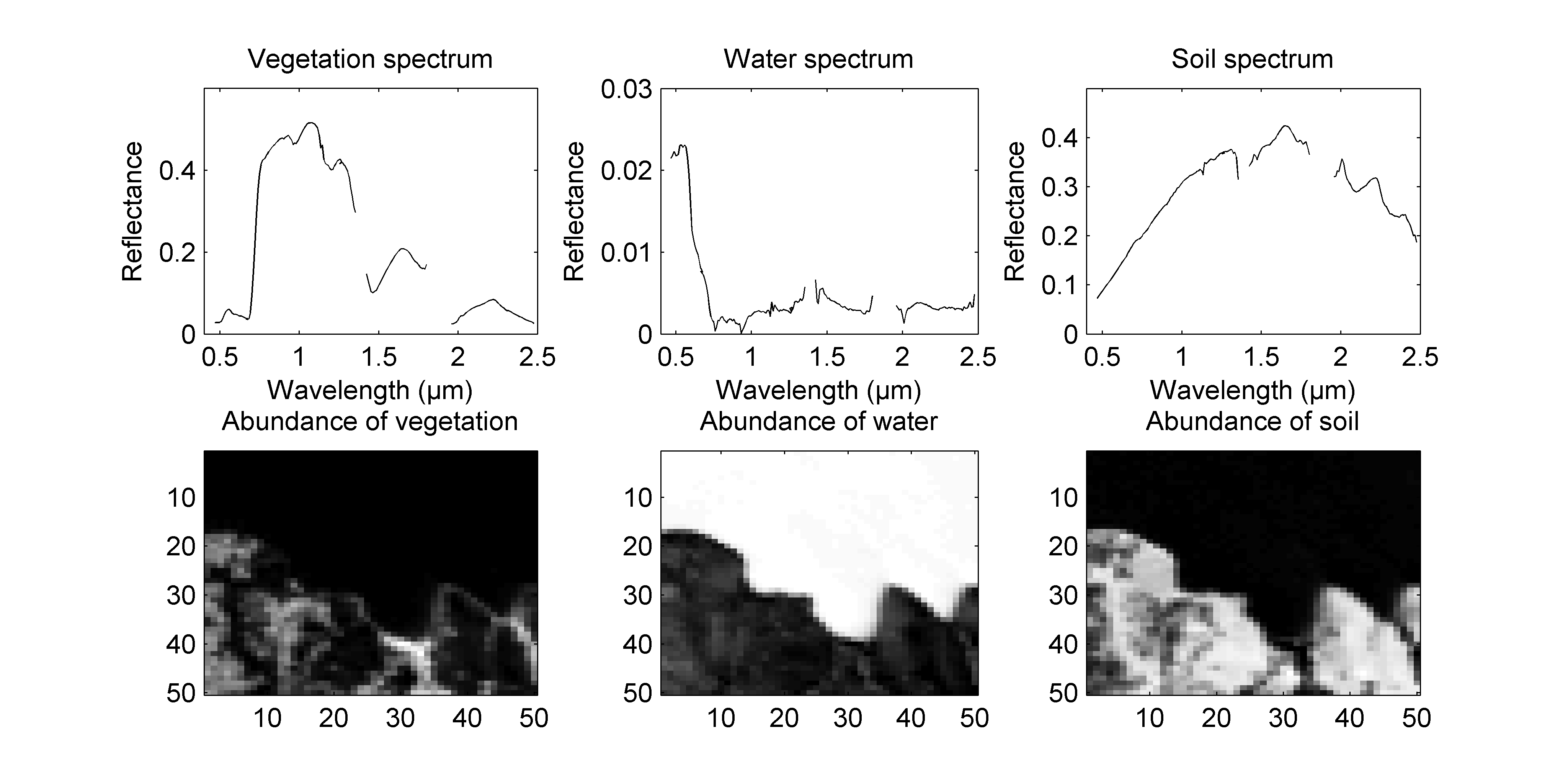}
  \caption{Top: the $\nbmat=3$ endmember spectra estimated by the algorithm. Bottom: the corresponding abundance maps
  (black (respectively, white) means absence (respectively, presence) of the
material).}\label{fig:Moffett_results}
\end{figure*}

A $50 \times 50$ sub-image depicted in Fig.~\ref{fig:Moffett_ROI}
(right) has been unmixed using the proposed Bayesian approach. The
number of endmembers has been estimated as in \cite{Dobigeon2008}.
More precisely, we retain the first $\nbmat-1$ eigenvalues
identified by PCA that capture $95\%$ of the energy contained into
the dataset. As detailed in \ref{subsubsec:dim_reduction}, we use
also PCA to choose the subset $\calV_{\nbmat-1}$ defined in
\eqref{eq:Vspace}. After a short burn-in period
$N_{\textrm{bi}}=50$, estimates of the parameters of interest are
computing following the MMSE principle in \eqref{eq:MMSE} with
$N_r=450$. The $\nbmat=3$ endmembers recovered by the proposed joint
Bayesian LSMA algorithm are depicted in
Fig.~\ref{fig:Moffett_results} (top). These endmember spectra are
represented in $\nbband=189$ spectral bands after removing the water
absorption bands\footnote{The water vapor absorption bands are
usually discarded to avoid poor SNR in these intervals.}. These
endmembers are characteristic of the coastal area that appears in
the image: vegetation, water and soil. The corresponding abundance
maps, shown Fig.~\ref{fig:Moffett_results} (bottom), are in
agreement with the previous results presented in
\cite{Dobigeon2008}.

\section{Conclusions}
\label{sec:conclusions} This paper presented a Bayesian model as
well as an MCMC algorithm for unsupervised unmixing of hyperspectral
images, i.e. estimating the endmember spectra in the observed scene
and their respective abundances for each pixel. Appropriate priors
were chosen for the abundance vectors to ensure non-negativity and
sum-to-one constraints inherent to the linear mixing model. Instead
of estimating the endmember spectral signatures in the observation
space, we proposed to estimate their projections onto a suitable
subspace. In this subspace, which can be identified by a standard
dimension reduction technique such as PCA, MNF and HySime, these
projections were assigned priors that satisfy positivity constraints
of the reconstructed endmember spectra. Due to the complexity of the
posterior distribution, a Gibbs sampling scheme was proposed to
generate samples asymptotically distributed according to this
posterior. The available samples were then used to approximate the
Bayesian estimators for the different parameters of interest.
Results of simulations conducted on synthetic and real hyperspectral
image illustrated the accuracy of the proposed Bayesian method when
compared with other algorithms from the literature. An interesting
open question is whether one can improve performance further by
folding the intrinsic dimension $\nbpband$ of the projection
subspace $\calV_\nbpband$ into the Bayesian framework, e.g., by
applying Bayesian PCA or Bayesian latent variable models. This
question is a topic of  our current research. While this paper
introduced a Bayesian method in the context of hyperspectral
unmixing, the method can also be used for other unmixing
applications, such as blind source separation, that satisfy
positivity and sum-to-one constraints.

\label{app:optimal_simplex}
\appendix[On the choice of uniform distributions as prior
distributions for $\Vabond{\nomat}$ and the size of the simplex
solution of the BSS problem] \label{app:optimal_simplex}

In this appendix, we show that choosing uniform distributions as
priors for the abundance vectors allows one to favor \emph{a
posteriori}, among two \emph{a priori} equiprobable polytopes that
are admissible solutions of the BSS problem, the solution
corresponding to the smallest polytope.\\

\emph{Property}: Let $\MATmat^{(1)}$ and $\MATmat^{(2)}$ be two
$\nbmat$-dimensional convex polytopes of $\R^{\nbband}$ that are
admissible solutions of the BSS constrained problem, i.e.
\begin{equation}
  \begin{array}{c}
    \exists \MATabond^{(1)}=\left[\Vabond{1}^{(1)},\ldots,\Vabond{\nbmat}^{(1)}\right]\transp \subset \calA^{\nbmat},  \\
    \exists \MATabond^{(2)}=\left[\Vabond{1}^{(2)},\ldots,\Vabond{\nbmat}^{(2)}\right]\transp \subset \calA^{\nbmat},  \\
  \end{array}
\end{equation}
such as
$\MATpix=\MATmat^{(1)}\MATabond^{(1)}=\MATmat^{(2)}\MATabond^{(2)}$
where $\calA$ has been defined in \eqref{eq:Aspace}. Then
\begin{equation}\label{eq:property}
\begin{split}
   &f\left({\MATmat}^{(1)}|\MATpix\right) \geq   f\left({\MATmat}^{(2)}|\MATpix\right)\\
    \Leftrightarrow&\\
   &\textrm{vol}\left(\Simplex_{{\MATmat}^{(1)}}\right) \leq
    \textrm{vol}\left(\Simplex_{{\MATmat}^{(2)}}\right),
\end{split}
\end{equation}
where $\textrm{vol}\left(\Simplex_{{\MATmat}^{(i)}}\right)$ stands
for the volume of the polytope $\Simplex_{{\MATmat}^{(i)}} \subset
\R^{\nbband}$ introduced in \eqref{eq:polytope_S} whose vertices are
the columns of $\MATmat^{(i)}$.\\

\emph{Proof}: First note that, in absence of noise, as
$\Vpix{\nopix} = \MATmat\Vabond{\nopix}$,
\begin{equation}
  \Vabond{\nopix} \sim \calU\left(\calA\right) \Leftrightarrow
\Vpix{\nopix}|\MATmat \sim \calU\left(\Simplex_{\MATmat}\right),
\end{equation}
where $\calU\left(\cdot\right)$ stands for the uniform distribution.

 Consequently,
\begin{equation}
  f\left(\MATpix|\MATmat\right) =
  \left[\frac{1}{\mathrm{vol}\left(\Simplex_{\MATmat}\right)}\right]^{\nbpix}
  \prod_{\nopix}^{\nbpix}\Indicfun{\Simplex_{\MATmat}}{\Vpix{\nopix}},
\end{equation}
which can be simplified by
\begin{equation}
\label{eq:YcondM}
  f\left(\MATpix|\MATmat\right) =
  \left[\frac{1}{\mathrm{vol}\left(\Simplex_{\MATmat}\right)}\right]^{\nbpix},
\end{equation}
since, by definition, the observed pixels $\Vpix{\nopix}$
($\nopix=1,\ldots,\nbpix$) belong to the solution polytope
$\Simplex_{\MATmat}$. Moreover, Bayes' paradigm allows one to state:
\begin{equation}
  f\left(\MATmat|\MATpix\right)  =
  \frac{f\left(\MATpix|\MATmat\right)f\left(\MATmat\right)}{f\left(\MATpix\right)}.
\end{equation}
Since the two solutions ${\MATmat^{(1)}}$ and ${\MATmat^{(2)}}$ are
\emph{a priori} equiprobable, from \eqref{eq:YcondM}, it yields:
\begin{equation}
  \frac{f\left({\MATmat^{(1)}}|\MATpix\right)}{f\left({\MATmat^{(2)}}|\MATpix\right)}
  =\left[\frac{\mathrm{vol}\left(\Simplex_{{\MATmat}^{(2)}}\right)}{\mathrm{vol}\left(\Simplex_{{\MATmat}^{(1)}}\right)}\right]^{\nbpix}.
\end{equation}
It follows \eqref{eq:property}.
\begin{flushright}
  \small{$\blacksquare$}
\end{flushright}
Note that the equiprobability assumption underlying the solutions
$\Simplex_{{\MATmat}^{(2)}}$ and $\Simplex_{{\MATmat}^{(2)}}$ is not
a too restrictive hypothesis. Indeed if the variances $s^2_\nomat$
($\nomat=1,\ldots,\nbmat$) had been chosen such that the prior
distribution in \eqref{eq:pmat_prior} is sufficiently flat, then:
\begin{equation}
  f\left(\MATmat^{(1)}\right) \approx f\left(\MATmat^{(2)}\right).
\end{equation}
Note also that the projection of the polytope
$\Simplex_{{\MATmat}^{(i)}}$ onto the subset
$\calV_{\nbmat-1}\subset \R^{\nbmat-1}$ is the simplex
$\Simplex_{{\MATpmat}^{(i)}}$ whose vertices are the columns of
$\MATpmat^{(i)}$.

\section*{Acknowledgments}
\label{sec:Ack} The authors would like to thank J. Idier and E. Le
Carpentier, from IRCCyN Nantes, for interesting discussions
regarding this work.

\bibliographystyle{ieeetran}

\begin{thebibliography}{10}
\providecommand{\url}[1]{#1} \csname url@samestyle\endcsname
\providecommand{\newblock}{\relax} \providecommand{\bibinfo}[2]{#2}
\providecommand{\BIBentrySTDinterwordspacing}{\spaceskip=0pt\relax}
\providecommand{\BIBentryALTinterwordstretchfactor}{4}
\providecommand{\BIBentryALTinterwordspacing}{\spaceskip=\fontdimen2\font
plus \BIBentryALTinterwordstretchfactor\fontdimen3\font minus
  \fontdimen4\font\relax}
\providecommand{\BIBforeignlanguage}[2]{{%
\expandafter\ifx\csname l@#1\endcsname\relax
\typeout{** WARNING: IEEEtran.bst: No hyphenation pattern has been}%
\typeout{** loaded for the language `#1'. Using the pattern for}%
\typeout{** the default language instead.}%
\else \language=\csname l@#1\endcsname \fi #2}}
\providecommand{\BIBdecl}{\relax} \BIBdecl

\bibitem{Chang2003}
{C.-I Chang}, \emph{Hyperspectral Imaging: Techniques for Spectral
detection
  and classification}.\hskip 1em plus 0.5em minus 0.4em\relax New York: Kluwer,
  2003.

\bibitem{Plaza2005}
J.~Plaza, R.~P\'erez, A.~Plaza, P.~Mart\'inez, and D.~Valencia,
``Mapping oil
  spills on sea water using spectral mixture analysis of hyperspectral image
  data,'' in \emph{Chemical and Biological Standoff Detection III}, J.~O.
  Jensen and J.-M. Th\'eriault, Eds., vol. 5995.\hskip 1em plus 0.5em minus
  0.4em\relax SPIE, 2005, pp. 79--86.

\bibitem{Manolakis2001}
D.~Manolakis, C.~Siracusa, and G.~Shaw, ``Hyperspectral subpixel
target
  detection using the linear mixing model,'' \emph{IEEE Trans. Geosci. and
  Remote Sensing}, vol.~39, no.~7, pp. 1392--1409, July 2001.

\bibitem{Keshava2002}
N.~Keshava and J.~F. Mustard, ``Spectral unmixing,'' \emph{IEEE
Signal
  Processing Magazine}, pp. 44--–57, Jan. 2002.

\bibitem{Singer1979}
R.~B. Singer and T.~B. McCord, ``Mars: Large scale mixing of bright
and dark
  surface materials and implications for analysis of spectral reflectance,'' in
  \emph{Proc. 10th Lunar and Planetary Sci. Conf.}, March 1979, pp. 1835--1848.

\bibitem{Hapke1981}
B.~W. Hapke, ``Bidirectional reflectance spectroscopy. {I}.
{T}heory,''
  \emph{J. Geophys. Res.}, vol.~86, pp. 3039--–3054, 1981.

\bibitem{Johnson1983}
P.~E. Johnson, M.~O. Smith, S.~Taylor-George, and J.~B. Adams, ``A
  semiempirical method for analysis of the reflectance spectra of binary
  mineral mixtures,'' \emph{J. Geophys. Res.}, vol.~88, pp. 3557--3561, 1983.

\bibitem{Boardman1993}
J.~Boardman, ``Automating spectral unmixing of {AVIRIS} data using
convex
  geometry concepts,'' in \emph{Summaries 4th Annu. {JPL} Airborne Geoscience
  Workshop}, vol.~1.\hskip 1em plus 0.5em minus 0.4em\relax Washington, D.C.:
  JPL Pub., 1993, pp. 11--14.

\bibitem{Winter1999}
M.~Winter, ``Fast autonomous spectral end-member determination in
hyperspectral
  data,'' in \emph{Proc. 13th Int. Conf. on Applied Geologic Remote Sensing},
  vol.~2, Vancouver, April 1999, pp. 337--344.

\bibitem{Nascimento2005}
J.~M. Nascimento and J.~M. {Bioucas-Dias}, ``Vertex component
analysis: A fast
  algorithm to unmix hyperspectral data,'' \emph{IEEE Trans. Geosci. and Remote
  Sensing}, vol.~43, no.~4, pp. 898--910, April 2005.

\bibitem{Craig1994}
M.~Craig, ``Minimum volume transforms for remotely sensed data,''
\emph{IEEE
  Trans. Geosci. and Remote Sensing}, pp. 542--552, 1994.

\bibitem{Bowles1995}
J.~H. Bowles, P.~J. Palmadesso, J.~A. Antoniades, M.~M. Baumback,
and L.~J.
  Rickard, ``Use of filter vectors and fast convex set methods in hyperspectral
  analysis,'' in \emph{Infrared Spaceborne Remote Sensing III}, M.~Strojnik and
  B.~F. Andresen, Eds., vol. 2553, no. 148.\hskip 1em plus 0.5em minus
  0.4em\relax SPIE, Sept. 1995, pp. 148--157.

\bibitem{Plaza2004}
A.~Plaza, P.~Martinez, R.~Perez, and J.~Plaza, ``A quantitative and
comparative
  analysis of endmember extraction algorithms from hyperspectral data,''
  \emph{IEEE Trans. Geosci. and Remote Sensing}, vol.~42, no.~3, pp. 650--663,
  March 2004.

\bibitem{Martinez2006}
P.~J. Martinez, R.~M. P\'erez, A.~Plaza, P.~L. Aguilar, M.~C.
Cantero, and
  J.~Plaza, ``Endmember extraction algorithms from hyperspectral images,''
  \emph{Annals of Geophysics}, vol.~49, no.~1, pp. 93--101, Feb. 2006.

\bibitem{Keshava2000}
N.~Keshava, J.~P. Kerekes, D.~G. Manolakis, and G.~A. Shaw,
``Algorithm
  taxonomy for hyperspectral unmixing,'' in \emph{Algorithms for Multispectral,
  Hyperspectral, and Ultraspectral Imagery VI}, S.~S. Shen and M.~R. Descour,
  Eds., vol. 4049, no.~1.\hskip 1em plus 0.5em minus 0.4em\relax SPIE, 2000,
  pp. 42--63.

\bibitem{Heinz2001}
D.~C. Heinz and {C.-I Chang}, ``Fully constrained least-squares
linear spectral
  mixture analysis method for material quantification in hyperspectral
  imagery,'' \emph{IEEE Trans. Geosci. and Remote Sensing}, vol.~29, no.~3, pp.
  529--545, March 2001.

\bibitem{Settle1996}
J.~Settle, ``On the relationship between spectral unmixing and
subspace
  projection,'' \emph{IEEE Trans. Geosci. and Remote Sensing}, vol.~34, no.~4,
  pp. 1045--1046, July 1996.

\bibitem{Dobigeon2008}
N.~Dobigeon, J.-Y. Tourneret, and {C.-I Chang}, ``Semi-supervised
linear
  spectral unmixing using a hierarchical {B}ayesian model for hyperspectral
  imagery,'' \emph{IEEE Trans. Signal Processing}, vol.~56, no.~7, pp.
  2684--2695, July 2008.

\bibitem{Comon1991}
P.~Common, C.~Jutten, and J.~Herault, ``Blind separation of sources.
{P}art
  {II}: problems statement,'' \emph{Signal Processing}, vol.~24, no.~1, pp.
  11--20, July 1991.

\bibitem{Lee1998}
T.~W. Lee, \emph{Independent component analysis: theory and
  applications}.\hskip 1em plus 0.5em minus 0.4em\relax Hingham, MA: Kluwer
  Academic Publishers, 1998.

\bibitem{Bayliss1998}
J.~Bayliss, J.~A. Gualtieri, and R.~Cromp, ``Analyzing hyperspectral
data with
  independent component analysis,'' in \emph{Proc. AIPR Workshop Exploiting New
  Image Sources and Sensors}, J.~M. Selander, Ed., vol. 3240.\hskip 1em plus
  0.5em minus 0.4em\relax Washington, D.C.: SPIE, 1998, pp. 133--143.

\bibitem{Dobigeon_SPIE_ERS_2005}
N.~Dobigeon and V.~Achard, ``Performance comparison of geometric and
  statistical methods for endmembers extraction in hyperspectral imagery,'' in
  \emph{Image and Signal Processing for Remote Sensing XI}, L.~Bruzzone, Ed.,
  vol. 5982, no.~1.\hskip 1em plus 0.5em minus 0.4em\relax SPIE, Oct. 2005, pp.
  335--344.

\bibitem{Nascimento2005b}
J.~M.~P. Nascimento and J.~M. {Bioucas-Dias}, ``Does independent
component
  analysis play a role in unmixing hyperspectral data?'' \emph{IEEE Trans.
  Geosci. and Remote Sensing}, vol.~43, no.~1, pp. 175--187, Jan. 2005.

\bibitem{Nascimento2007igarss}
J.~M. Nascimento and J.~M. {Bioucas-Dias}, ``Hyperspectral unmixing
algorithm
  via dependent component analysis,'' vol.~43, no.~4, July 2007, pp.
  4033--4036.

\bibitem{Paatero1994}
P.~Paatero and U.~Tapper, ``Positive matrix factorization: a
non-negative
  factor model with optimal utilization of error estimates of data values,''
  \emph{Environmetrics}, vol.~5, pp. 111--126, 1994.

\bibitem{Sajda2004}
P.~Sajda, S.~Du, T.~R. Brown, R.~Stoyanova, D.~C. Shungu, X.~Mao,
and L.~C.
  Parra, ``Nonnegative matrix factorization for rapid recovery of constituent
  spectra in magnetic resonance chemical shift imaging of the brain,''
  \emph{IEEE Trans. Medical Imaging}, vol.~23, no.~12, pp. 1453--1465, 2004.

\bibitem{Berman2004}
M.~Berman, H.~Kiiveri, R.~Lagerstrom, A.~Ernst, R.~Dunne, and J.~F.
Huntington,
  ``{ICE}: A statistical approach to identifying endmembers in hyperspectral
  images,'' \emph{IEEE Trans. Geosci. and Remote Sensing}, vol.~42, no.~10, pp.
  2085--2095, Oct. 2004.

\bibitem{Miao2007}
L.~Miao and H.~Qi, ``Endmember extraction from highly mixed data
using minimum
  volume constrained nonnegative matrix factorization,'' \emph{IEEE Trans.
  Geosci. and Remote Sensing}, vol.~45, no.~3, pp. 765--–777, March 2007.

\bibitem{Rowe2002}
D.~B. Rowe, ``A {B}ayesian approach to blind source separation,''
\emph{J. of
  Interdisciplinary Mathematics}, vol.~5, no.~1, pp. 49--76, 2002.

\bibitem{Fevotte2006}
C.~F\'evotte and S.~J. Godsill, ``A {B}ayesian approach for blind
separation of
  sparse sources,'' \emph{IEEE Trans. Audio, Speech, Language Processing},
  vol.~14, no.~6, pp. 2174--2188, Nov. 2006.

\bibitem{Moussaoui2006}
S.~Moussaoui, D.~Brie, A.~{Mohammad-Djafari}, and C.~Carteret,
``Separation of
  non-negative mixture of non-negative sources using a {B}ayesian approach and
  {MCMC} sampling,'' \emph{IEEE Trans. Signal Processing}, vol.~54, no.~11, pp.
  4133--4145, Nov. 2006.

\bibitem{Dobigeon2007ssp}
N.~Dobigeon, S.~Moussaoui, and J.-Y. Tourneret, ``Blind unmixing of
linear
  mixtures using a hierarchical bayesian model. {A}pplication to spectroscopic
  signal analysis,'' in \emph{Proc. IEEE-SP Workshop Stat. and Signal
  Processing}, Madison, USA, Aug. 2007, pp. 79--83.

\bibitem{Harsanyi1994}
J.~C. Harsanyi and {C.-I Chang}, ``Hyperspectral image
classification and
  dimensionality reduction: An orthogonal subspace projection approach,''
  \emph{IEEE Trans. Geosci. and Remote Sensing}, vol.~32, no.~4, pp. 779--785,
  July 1994.

\bibitem{Chang1998}
{C.-I Chang}, ``Further results on relationship between spectral
unmixing and
  subspace projection,'' \emph{IEEE Trans. Geosci. and Remote Sensing},
  vol.~36, no.~3, pp. 1030--1032, May 1998.

\bibitem{Chang1998b}
{C.-I Chang}, X.-L. Zhao, M.~L.~G. Althouse, and J.~J. Pan, ``Least
squares
  subspace projection approach to mixed pixel classification for hyperspectral
  images,'' \emph{IEEE Trans. Geosci. and Remote Sensing}, vol.~36, no.~3, pp.
  898--912, May 1998.

\bibitem{DobigeonTR2008b}
\BIBentryALTinterwordspacing N.~Dobigeon and J.-Y. Tourneret,
``Bayesian sampling of structured noise
  covariance matrix for hyperspectral imagery,'' University of Toulouse, Tech.
  Rep., Dec. 2008. [Online]. Available:
  \url{http://dobigeon.perso.enseeiht.fr/publis.html}
\BIBentrySTDinterwordspacing

\bibitem{Li2004}
J.~Li, ``Wavelet-based feature extraction for improved endmember
abundance
  estimation in linear unmixing of hyperspectral signals,'' \emph{IEEE Trans.
  Geosci. and Remote Sensing}, vol.~42, no.~3, pp. 644--649, March 2004.

\bibitem{Chang2006}
{C.-I Chang} and B.~Ji, ``Weighted abundance-constrained linear
spectral
  mixture analysis,'' \emph{IEEE Trans. Geosci. and Remote Sensing}, vol.~44,
  no.~2, pp. 378--388, Feb. 2001.

\bibitem{Veit2009}
T.~Veit, J.~Idier, and S.~Moussaoui, ``{R}\'e\'echantillonnage de
l'\'echelle
  dans les algorithmes {MCMC} pour les probl\`emes inverses bilin\'eaires,''
  \emph{{{T}raitement du {S}ignal}}, 2009, to appear.

\bibitem{Moussaoui2005icassp}
S.~Moussaoui, D.~Brie, and J.~Idier, ``Non-negative source
separation: range of
  admissible solutions and conditions for the uniqueness of the solution,'' in
  \emph{Proc. IEEE Int. Conf. Acoust., Speech, and Signal Processing (ICASSP)},
  vol.~5, Philadelphia, USA, March 2005, pp. 289--292.

\bibitem{Jolliffe1986}
I.~T. Jolliffe, \emph{Principal Component Analysis}.\hskip 1em plus
0.5em minus
  0.4em\relax New York: Springer-Verlag, 1986.

\bibitem{Green1988}
A.~A. Green, M.~Berman, P.~Switzer, and M.~D. Craig, ``A
transformation for
  ordering multispectral data in terms of image quality with implications for
  noise removal,'' \emph{IEEE Trans. Geosci. and Remote Sensing}, vol.~26,
  no.~1, pp. 65--74, Jan. 1988.

\bibitem{Robert2007}
C.~P. Robert, \emph{The Bayesian Choice: from Decision-Theoretic
Motivations to
  Computational Implementation}, 2nd~ed., ser. Springer Texts in
  Statistics.\hskip 1em plus 0.5em minus 0.4em\relax New York: Springer-Verlag,
  2007.

\bibitem{Punskaya2002}
E.~Punskaya, C.~Andrieu, A.~Doucet, and W.~Fitzgerald, ``{B}ayesian
curve
  fitting using {M}{C}{M}{C} with applications to signal segmentation,''
  \emph{IEEE Trans. Signal Processing}, vol.~50, no.~3, pp. 747--758, March
  2002.

\bibitem{Dobigeon2007b}
N.~Dobigeon, J.-Y. Tourneret, and M.~Davy, ``Joint segmentation of
piecewise
  constant autoregressive processes by using a hierarchical model and a
  {B}ayesian sampling approach,'' \emph{IEEE Trans. Signal Processing},
  vol.~55, no.~4, pp. 1251--1263, April 2007.

\bibitem{Jeffreys1961}
H.~Jeffreys, \emph{Theory of Probability}, 3rd~ed.\hskip 1em plus
0.5em minus
  0.4em\relax London: Oxford University Press, 1961.

\bibitem{Robert1999}
C.~P. Robert and G.~Casella, \emph{Monte Carlo Statistical
Methods}.\hskip 1em
  plus 0.5em minus 0.4em\relax New York: Springer-Verlag, 1999.

\bibitem{DobigeonTR2007b}
\BIBentryALTinterwordspacing N.~Dobigeon and J.-Y. Tourneret,
``Efficient sampling according to a
  multivariate {G}aussian distribution truncated on a simplex,''
  IRIT/ENSEEIHT/T\'eSA, Tech. Rep., March 2007. [Online]. Available:
  \url{http://dobigeon.perso.enseeiht.fr/publis.html}
\BIBentrySTDinterwordspacing

\bibitem{Kay1988}
S.~M. Kay, \emph{Fundamentals of Statistical Signal Processing:
Estimation
  theory}.\hskip 1em plus 0.5em minus 0.4em\relax Englewood Cliffs NJ: Prentice
  Hall, 1993.

\bibitem{Robert1995}
C.~P. Robert, ``Simulation of truncated normal variables,''
\emph{Statistics
  and Computing}, vol.~5, pp. 121--125, 1995.

\bibitem{Devroye1986}
L.~Devroye, \emph{Non-Uniform Random Variate Generation}.\hskip 1em
plus 0.5em
  minus 0.4em\relax New York: Springer-Verlag, 1986.

\bibitem{ENVImanual2003}
{{RSI} (Research Systems Inc.)}, \emph{ENVI User's guide Version
4.0}, Boulder,
  CO 80301 USA, Sept. 2003.

\bibitem{Bioucas2008}
J.~M. {Bioucas-Dias} and J.~M.~P. Nascimento, ``Hyperspectral
subspace
  identification,'' \emph{IEEE Trans. Geosci. and Remote Sensing}, vol.~46,
  no.~8, pp. 2435--2445, Aug. 2008.

\bibitem{JPL_AVIRIS}
\BIBentryALTinterwordspacing {{AVIRIS} Free Data}. (2006) Jet
Propulsion Lab. (JPL). California Inst.
  Technol., Pasadena, CA. [Online]. Available:
  \url{http://aviris.jpl.nasa.gov/html/aviris.freedata.html}
\BIBentrySTDinterwordspacing

\bibitem{Christophe2005}
E.~Christophe, D.~L\'eger, and C.~Mailhes, ``Quality criteria
benchmark for
  hyperspectral imagery,'' \emph{IEEE Trans. Geosci. and Remote Sensing},
  vol.~43, no.~9, pp. 2103--2114, Sept. 2005.

\bibitem{Akgun2005}
T.~Akgun, Y.~Altunbasak, and R.~M. Mersereau, ``Super-resolution
reconstruction
  of hyperspectral images,'' \emph{IEEE Trans. Image Processing}, vol.~14,
  no.~11, pp. 1860--1875, Nov. 2005.

\end{thebibliography}

\end{document}